%% file: main.tex
\documentclass[11pt]{article}
\usepackage[final]{acl}
\AtBeginDocument{%
  }

\usepackage{times}
\usepackage{latexsym}
\usepackage[T1]{fontenc}
\usepackage[utf8]{inputenc}
\usepackage{microtype}
\usepackage{amssymb}
\usepackage{array}
\usepackage{booktabs}
\usepackage{tabularx}
\usepackage{makecell}
\usepackage{comment}
\usepackage{graphicx} 
\usepackage{caption}  
\usepackage{subcaption} 

\usepackage[autolanguage, np]{numprint}
\usepackage{multirow}
\usepackage{graphicx}
\usepackage{tabularx}
\usepackage{booktabs}
\usepackage{multirow}
\usepackage{amsmath, mathrsfs}

\renewcommand\arraystretch{1.3} 
\newcolumntype{Z}{>{\centering\arraybackslash}X}

\usepackage{graphicx}
\usepackage[table]{xcolor}
\usepackage{booktabs}

\usepackage[T1]{fontenc}
\usepackage{fontawesome5}
\usepackage{pifont}
\usepackage{xcolor}

\newcommand{\best}{\cellcolor{teal!25}}

\newcommand{\sysname}{\texttt{RespiraMFM}~}
\begin{document}

\title{
\raisebox{-0.35\height}{\includegraphics[height=2.0em]{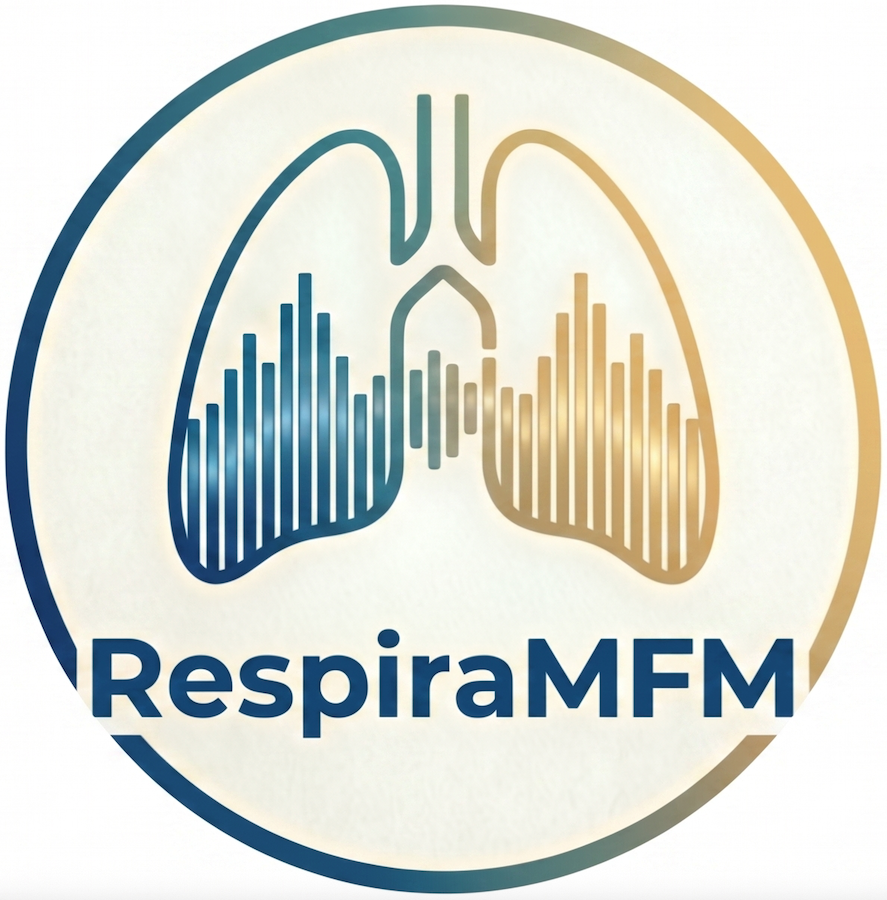}}~RespiraMFM: A Multimodal Foundation Model with Contrastive Audio-Language Alignment for Respiratory Disease Identification
}


\author{
Shakhrul Iman Siam\textsuperscript{1},
Tiantian Feng\textsuperscript{2},
Jiankun Zhang\textsuperscript{3},\\
\textbf{Shrikanth Narayanan}\textsuperscript{2},
\textbf{Mi Zhang}\textsuperscript{1} \\
\textsuperscript{1}The Ohio State University \quad
\textsuperscript{2} University of Southern California \quad
\textsuperscript{3}University of Chicago \quad \\
\texttt{\{siam.5, mizhang.1\}@osu.edu} \\
\href{https://respiramfm.github.io/}{\faGlobe~Project Page} \quad 
\href{https://github.com/AIoT-MLSys-Lab/RespiraMFM}{\faGithub~GitHub}
}

\maketitle

\begin{abstract}
\input{sections/0_Abstract}
\end{abstract}

\input{sections/1_Introduction}

\input{sections/2_Related_work}
\input{sections/3_Methodology}

\input{sections/4_Experiments}

\input{sections/5_Results}

\input{sections/6_Conclusion}

\input{sections/7_Limitations}

\input{sections/8_Ethical}

\bibliography{refs}

\appendix
\input{sections/Appendix}

\end{document}

%% file: sections/0_Abstract.tex
Respiratory diseases remain a leading cause of global mortality, where timely and accurate diagnosis is critical to improving patient outcomes and reducing healthcare burdens. While prior work has explored audio-based models for respiratory disease detection, such unimodal approaches often suffer from limited generalizability and diagnostic precision. 
In this paper, we propose \textbf{RespiraMFM}, a Multimodal Foundation Model that integrates respiratory sounds with patient medical history and symptoms to enhance diagnostic accuracy and disease detection capabilities.
We introduce an effective contrastive alignment strategy for audio-text multimodal integration, allowing the model to learn better cross-modal representations between respiratory sounds and corresponding textual clinical information.
We evaluate \texttt{RespiraMFM} across five major respiratory diseases using seven real-world datasets in both supervised fine-tuning and zero-shot settings, achieving a 9.15\% improvement in AUROC on supervised tasks and a 20.98\% gain on zero-shot tasks over existing baselines.
%
These findings underscore the potential of our framework to advance early diagnosis and improve clinical decision-making in respiratory disease management.
%

%% file: sections/1_Introduction.tex
\section{Introduction}

Respiratory diseases, such as COVID-19, tuberculosis (TB), chronic obstructive pulmonary disease (COPD), asthma, and pneumonia,  remain a leading cause of morbidity and mortality worldwide \cite{weinberger2020estimation}. 
Most existing works \cite{baur2024hear, zhang2024opera} on detecting those respiratory diseases rely solely on audio inputs such as coughing sounds or stethoscope recordings. 
However, their performance is often constrained by the limited information that audio data alone can provide. 

To mitigate the constraints of relying solely on audio inputs, multimodal methods \cite{kim2024bts, zhang2024respllm} that combine respiratory audio with relevant clinical information, such as symptoms (e.g., fever, fatigue, chest pain) and lifestyle factors such as smoking history, have been proposed.
For instance, BTS \cite{kim2024bts} concatenates learned representations from audio and text encoders to identify respiratory diseases.
%
RespLLM \cite{zhang2024respllm} utilizes a large language model (LLM) as the text encoder alongside a separate audio encoder, with a linear projector for matching dimensions.

%
%
Although combining audio input with text in a multi-modal setting has been widely adopted in recent works \cite{ma2024embarrassingly, zhang2025soundwave}, directly applying the existing approach in the respiratory disease detection task poses a unique challenge. Prior studies on audio–text alignment have primarily focused on spoken language, where audio carries rich semantic and syntactic information that can be naturally aligned with text. In contrast, the respiratory disease identification task involves fusing non-linguistic acoustic biomarkers (i.e., coughs, wheezes, and crackles) with free-form patient symptoms text.
Existing work on respiratory LLMs, such as RespLLM \cite{zhang2024respllm}, often adopts the same straightforward fusion techniques used for spoken audio, typically relying on simple feature concatenation or a trainable linear projector within a unified end-to-end framework. 
This strategy suffers from two critical limitations: First, because non-linguistic acoustic biomarkers like cough sounds are fundamentally misaligned with the text description of a patient's condition, the unified training objective struggles to converge on a stable shared representation, resulting in failure to leverage both modalities effectively. Second, due to this inefficient fusion mechanism being tightly coupled to the training data, these models only perform well on in-domain diseases and fail in zero-shot scenarios when encountering new, unseen diseases.
Our key technical contribution is identifying this misalignment problem and proposing a simple yet effective solution: an alignment stage that explicitly aligns respiratory audio embeddings with clinical text embeddings before fine-tuning.
We introduce \textbf{RespiraMFM}, a two-stage, decoupled training architecture for respiratory disease identification.
In our two-stage approach, the first stage is dedicated to Modality Alignment. A lightweight projector module is contrastively trained to learn a robust mapping that projects the high-dimensional audio features directly into the semantic embedding space of the LLM.
Reflecting the success of contrastive methods in improving zero-shot performance \cite{wu2023large}, this pre-training provides better initialization for the weights of the projector and ensures the acoustic features are semantically anchored to the correct symptom concepts.
For the second stage, we freeze the learned projector module, and the resulting aligned embeddings are then passed to the LLM to generate the final disease prediction.
We evaluate \sysname using seven real-world datasets that cover five of the most common respiratory diseases: COVID-19, TB, COPD, asthma, and pneumonia. We highlight four of our findings: 
(1) \sysname consistently outperforms state-of-the-art multimodal baselines on respiratory disease identification by achieving a 9.15\% improvement in AUROC on supervised tasks and a 20.98\% improvement on zero-shot tasks. 
(2)  \sysname achieves superior generalization capabilities and effectively detects unseen respiratory diseases without requiring any training samples of those diseases.
(3) \sysname significantly reduces the training data requirement, achieving comparable performance with an order of magnitude less training data compared to the baselines. 
(4) Our modality alignment module effectively unifies audio and text modalities, leading to consistent AUROC improvements across all tasks compared to models without this module.

%% file: sections/2_Related_work.tex
\section{Related Work}

\begin{figure*}[t]
    \centering
    \includegraphics[width=0.9\linewidth]{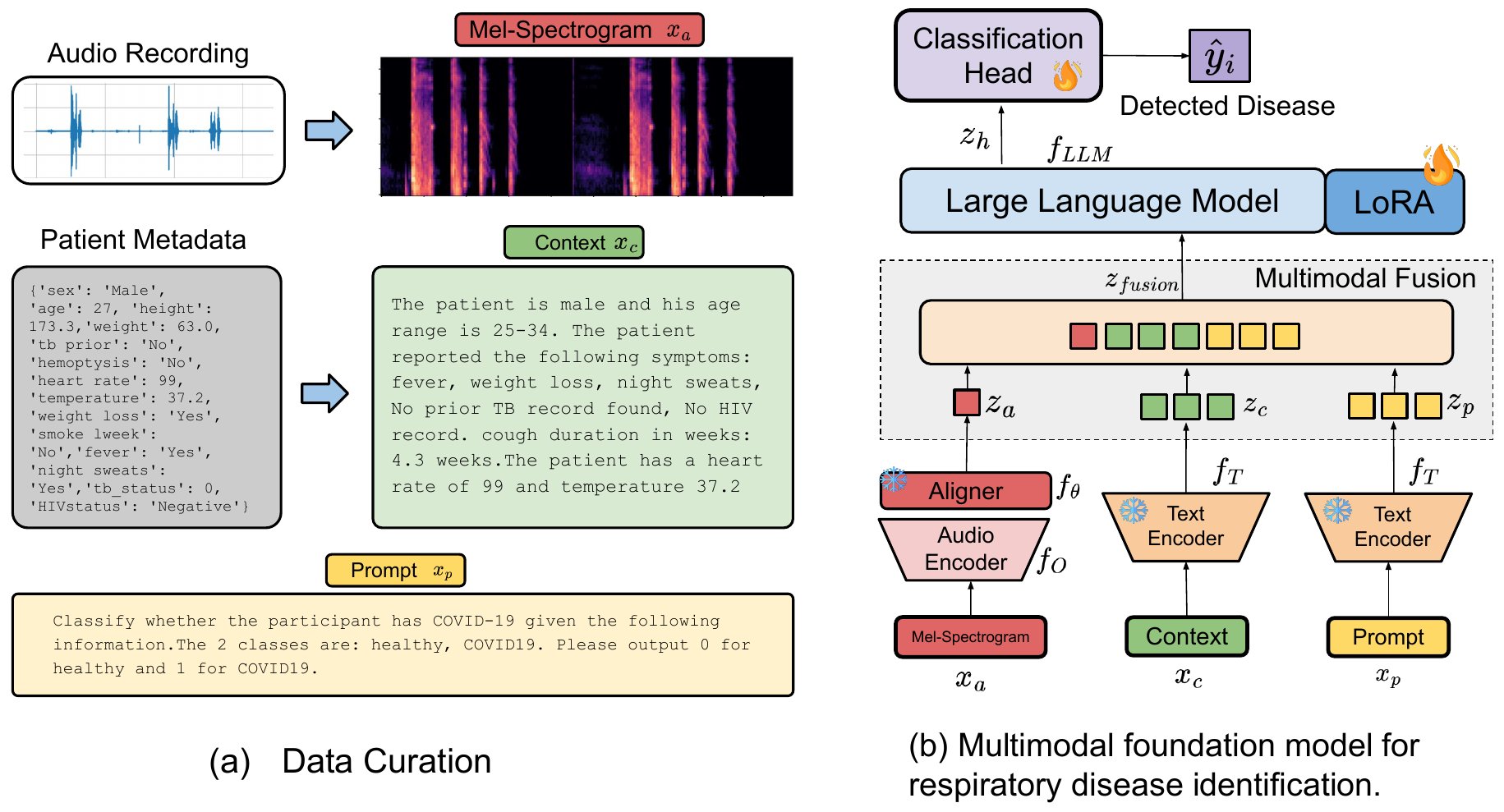}
    \vspace{-2mm}
    \caption{Overview of RespiraMFM.}
    \vspace{-3mm}
    \label{fig:model-architecture}
\end{figure*} 

\subsection{Single-Modal Models}
The majority of existing respiratory disease identification methods \cite{yang2020adventitious, ma2020lungrn+, chang2022example} rely solely on audio inputs such as cough sounds or stethoscope recordings. 
\citet{bae2023patch} introduces a contrastive learning framework to enhance respiratory sound classification using Audio Spectrogram Transformer (AST). By mixing spectrogram patches generated from raw audio data and applying contrastive loss, the model learns robust and discriminative features, which are subsequently passed to a linear classifier for respiratory disease identification.
OPERA \cite{zhang2024opera} curates large-scale unlabeled respiratory audio datasets and pretrains three foundational models using self-supervised learning. Among them, OPERA-CT, the best-performing model, is a contrastive learning–based transformer model, which is used as a general-purpose feature extractor for respiratory disease classification tasks. 
HeAR (Health Acoustic Representations) \cite{baur2024hear} introduces a self-supervised generative learning-based framework trained on a large corpus of health-related audio data. By leveraging generative objectives during pretraining, HeAR learns generalizable audio representations, which are utilized for downstream disease diagnosis tasks via simple linear probes. 
Despite the promising results of single-modal models, their performance is limited by the information available from audio data alone.

\subsection{Multimodal Models}
Unlike single-modal models, multimodal models combine audio data with textual information such as patient symptoms and medical history, leading to more accurate diagnoses.
BTS \cite{kim2024bts} introduces a text-audio model that combines respiratory sounds with metadata transformed into descriptive text. It uses the Contrastive Language-Audio Pretraining (CLAP) \cite{elizalde2023clap} model to extract features from both modalities, followed by a linear classifier for respiratory disease classification. However, the use of a basic linear classifier limits its ability to generalize in zero-shot scenarios or when encountering new diseases.
To date, RespLLM \cite{zhang2024respllm} is one of the early efforts that applies a multimodal LLM framework integrating text and audio representations for respiratory disease prediction. Their approach utilizes a pretrained encoder to extract audio and text features and a trainable linear projector to align the feature dimensions with LLM. 
However, since each modality encoder is trained separately, the resulting representations are often distinct and may not be directly compatible across modalities.
While a linear projector can align the encoder output dimensions with those expected by the LLM, it does not ensure semantic alignment between modalities. 
To address these limitations, we propose a contrastive alignment module that facilitates more effective integration by aligning audio and text representations in a shared semantic space. 
Our approach goes beyond mere dimensional alignment, aiming to establish a shared representation space that enables effective integration of multimodal information.

%% file: sections/3_Methodology.tex
\section{RespiraMFM}
\subsection{Overview}
Figure \ref{fig:model-architecture} provides an overview of the proposed \sysname framework. In the data curation stage shown in Figure \ref{fig:model-architecture}(a), given the multimodal respiratory datasets, we extract and pre-process the raw audio data and the corresponding patient symptoms to construct the instruction tuning data for respiratory disease identification.
%
%
As shown in Figure \ref{fig:model-architecture}(b),
the curated multimodal data are first processed by the audio and text encoders to get audio and text representations, respectively. One key component of our framework is the alignment module that reduces the domain mismatches between audio features and the language model embeddings. 
The alignment module is trained separately via contrastive learning.
Upon completion of training, the alignment module is frozen and incorporated into the instruction tuning stage.
During instruction tuning, the curated data are passed through each encoder to obtain modality-specific representations, which are then fused by concatenating them. The resulting multimodal representation is subsequently fed into the LLM to generate predictions for respiratory disease classification.
%

\subsection{Data Curation}
\label{section:data-curation} 
The multimodal respiratory datasets consist of both respiratory audio recordings and the corresponding patient-reported symptoms in either JSON or tabular format. 
The objective of data curation is to create the instruction tuning data by pre-processing the respiratory audio recordings, generating instruction prompts, and converting patient-reported symptoms into structured textual representations. 
For audio recordings, each recording was normalized to 8 seconds in length by either truncating longer recordings or padding shorter ones through repetition.
The audio signals are then processed with a 64ms Hann window with a 32ms step size, and subsequently converted into mel spectrograms denoted as $x_a$ using the features extracted from a pre-trained OPERA-CT encoder \cite{zhang2024opera}.
%
%
Patient metadata varies across datasets in terms of structure and format. As shown in Figure \ref{fig:model-architecture}(a), we select relevant symptoms (Table \ref{tab:selected-symptoms}) from the tabular data and apply a standardized template to generate a textual representation $x_c$. 
We utilize task-specific prompts $x_p$ such as -  
\textit{"Classify whether the participant has COVID-19 given the following information. The 2 classes are: healthy, COVID19. Please output 0 for healthy and 1 for COVID19"} to guide the LLM in producing disease classification outputs.


\subsection{Contrastive Learning-based Audio-Text Aligning}
\label{section:training-projector}
\begin{figure*}[]
    \centering
    \includegraphics[width=0.9\linewidth]{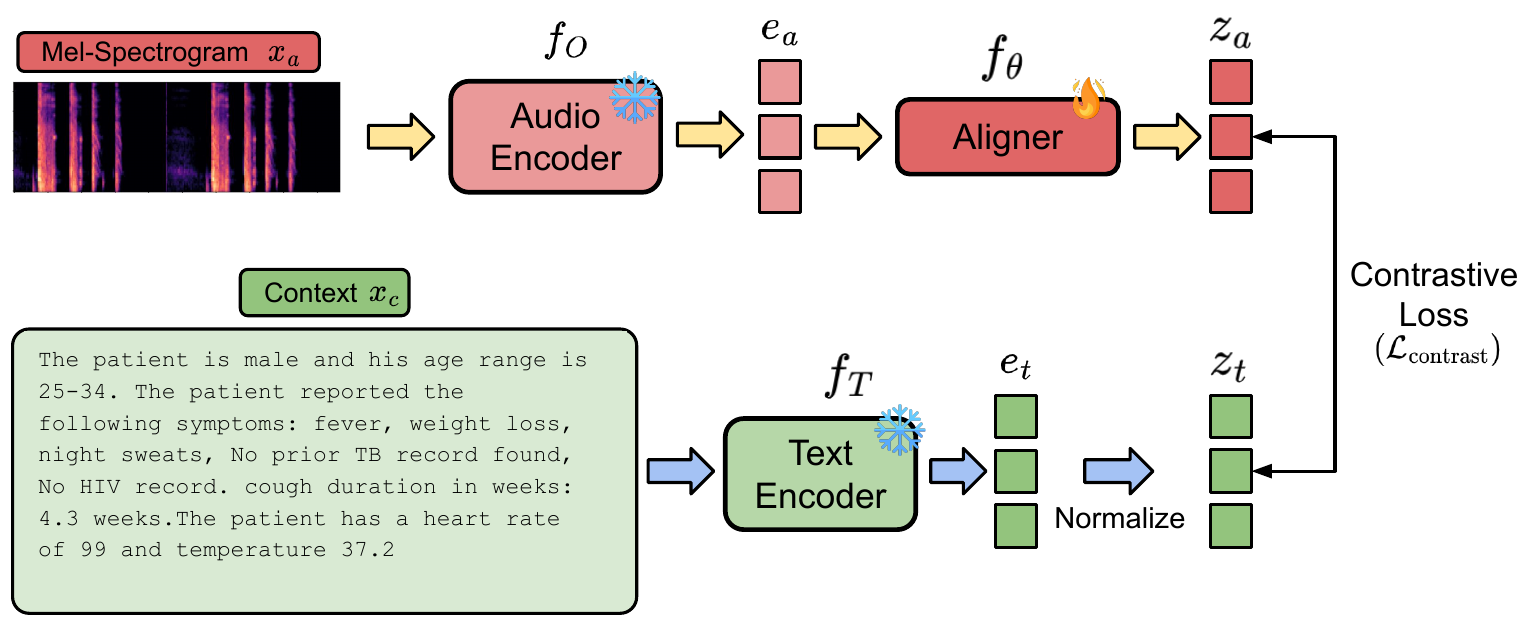}
    \caption{Illustration of contrastive learning-based audio-text alignment.}
    \label{fig:projector-contrastive}
\end{figure*}
We introduce a contrastive learning-based audio-text alignment module to align audio features with language model embeddings effectively. 
Specifically, we employ a pre-trained audio encoder that generates 768-dimensional embeddings from the input audio. In contrast, large language models typically operate with higher-dimensional input embeddings. Therefore, it is necessary to match the audio dimension to use as input into the LLM. 
Prior work \cite{zhang2024respllm} addresses this by introducing a trainable linear projector to map audio embeddings into the higher-dimensional space required by the LLM. 
However, given the fundamental differences between the audio and text encoders in both architecture and representational semantics, simple dimensional alignment may be insufficient to achieve effective multimodal fusion \cite{lyu2023macaw, lyu2024unibind}.
%
To enable more effective cross-modal alignment between audio and text, we adopt a contrastive learning strategy—an approach shown to yield powerful multimodal representations in models like CLIP \cite{radford2021learning}.
As shown in Figure \ref{fig:projector-contrastive}, we compute text embeddings \( \mathbf{e}_t \in \mathbb{R}^d \) using a frozen LLM, where \( \mathbf{e}_t = f_T(x_c) \), \( f_T \) is the text encoder (LLM), \( x_c \) is the textual context, and $d$ is the embedding dimension of LLM.
Similarly, audio embeddings \( \mathbf{e}_a \in \mathbb{R}^{768} \) are extracted using a frozen OPERA encoder, where  \( \mathbf{e}_a = f_O(x_a) \), \( f_O \) is the pre-trained Opera-CT audio encoder, and \( x_a \) is the mel-spectrogram of the raw audio data.

A lightweight projection head \(f_\theta: \mathbb{R}^{768} \rightarrow \mathbb{R}^d \) is trained to map audio embeddings into the same semantic space as the text embeddings. The training objective minimizes a contrastive loss that encourages matched audio-text pairs to be close while pushing unmatched pairs apart. 

Formally, for a batch of \( N \) paired samples, we define the normalized embeddings as:
\[
\mathbf{z}_i^a = \frac{f_\theta(\mathbf{e}_i^a)}{\|f_\theta(\mathbf{e}_i^a)\|}, \quad \mathbf{z}_i^t = \frac{\mathbf{e}_i^t}{\|\mathbf{e}_i^t\|}.
\]
The contrastive loss \cite{chen2020simple} is given by:
\[
\mathcal{L}_{\text{contrast}} = -\frac{1}{N} \sum_{i=1}^{N} \log \frac{\exp(\mathbf{z}_i^a \cdot \mathbf{z}_i^t / \tau)}{\sum_{j=1}^{N} \exp(\mathbf{z}_i^a \cdot \mathbf{z}_j^t / \tau)},
\]
where \( \tau \) is a temperature scaling factor. 
%
By training this projection head with contrastive supervision, we achieve better semantic alignment across modalities while keeping the audio and text encoders frozen. The model architecture and additional training details about the aligner module are presented in Appendix \ref{appendix: contrastive-aligner}.
%

\subsection{Instruction Tuning}
\label{section:instruction-tuning}
We employ instruction tuning to guide the LLM in understanding and following task-specific prompts that connect the multi-modal input and the corresponding diagnostic outcomes. The core components of instruction tuning are described below.

\vspace{1mm}
\noindent
\textbf{Multimodal Fusion:}
The audio and text features are fused at the embedding level by concatenation.
%
During this stage, we utilize the contrastively trained alignment module ($f_\theta$) from the previous step (\S \ref{section:training-projector}), keeping its weights frozen to preserve the learned representations.
Similarly, the text encoder ($f_T$) and the audio encoder ($f_O$) are also kept frozen during this stage. The inputs include mel-spectrogram ($x_a$), curated patient symptom descriptions as contextual information ($x_c$), and a task-specific prompt ($x_p$). Audio embeddings $z_a \in \mathbb{R}^d$ are extracted via the audio encoder and subsequently projected to match the input dimensionality of the LLM. 
\begin{align*}
    z_a = f_\theta(f_O(x_a))
\end{align*}

Simultaneously, the prompt and contextual text are processed through the LLM’s encoder to obtain their respective representations, denoted as $z_p \in \mathbb{R}^d$ and $z_c \in \mathbb{R}^d$, corresponding to the prompt and context embeddings.

\[
z_p = f_T(x_p), \quad z_c = f_T(x_c).
\]

Finally, we concatenate the audio $(z_a)$, prompt $(z_p)$, and context $(z_c)$ embeddings to get a combined embedding of a longer sequence:
\begin{align*}
    z_{fusion} &= z_a~\mathbin{\|}~z_p~\mathbin{\|}~z_c
\end{align*}
where $\mathbin{\|}$ denotes concatenation operation.

\vspace{1mm}
\noindent
\textbf{Large Language Model:}
We utilize Phi-2\footnote{https://huggingface.co/microsoft/phi-2}, a 2.7B parameter model, as the backbone LLM. 
To adapt it for our classification task, we extend the model by appending a linear classification head atop the transformer architecture. 
We first form the multimodal fusion embeddings by concatenating audio and text representations. These fused embeddings are then fed into the LLM to produce a sequence of hidden states. A pooling layer is then applied to obtain the latent representation $z_{h}$. Specifically, we adopt the default pooling strategy, which selects the hidden state corresponding to the final token in the sequence. Finally, a linear classification head is applied to the pooled representation to produce prediction scores for different respiratory disease identification tasks.


\begin{align*}
    z_{h} &= Pool_{final}(f_{LLM}(z_{fusion}))
\end{align*}

This vector $z_{h}$ is then passed through a classification head comprising fully connected layers, followed by a softmax function to produce class probability distributions. The model is trained using cross-entropy loss:

\[
\mathcal{L}_{\text{CE}} = -\sum_{i=1}^{C} y_i \log(\hat{y}_i)
\]
where \( y_i \) and \( \hat{y}_i \) are the true and predicted probabilities for class \( i \), respectively.

\noindent
\textbf{Training Details:}
The instruction tuning process combines task-specific instructions $x_p$ with multimodal audio ($x_a$) and text ($x_c$) inputs to ensure the model generates outputs that align with the desired response format.
Additionally, we use LoRA (Low-Rank Adaptation) \cite{hu2021lora}, a parameter-efficient fine-tuning (PEFT) technique designed to preserve the inherent knowledge of a pre-trained LLM. 
The model was fine-tuned for 20 epochs, and the training configuration further optimizes LoRA with parameters like a rank ($r$) of 16, scaling factor ($\alpha$) of 32, and a dropout of 0.1.

%% file: sections/4_Experiments.tex
\input{tables/dataset}

\section{Experimental Setup}
\subsection{Datasets and Tasks}
We evaluate the performance of \sysname using seven real-world datasets, covering five of the most common respiratory diseases: COVID-19, TB, COPD, asthma, and pneumonia. 
These datasets include both respiratory audio recordings (e.g., coughing sound, stethoscope sound) and the associated metadata, such as patient-reported symptoms and medical history. 
Based on these datasets, we construct nine respiratory disease identification tasks as summarized in Table \ref{tab:dataset-overview}. 
Datasets associated with tasks T1 through T4 are used for training and in-domain evaluation using held-out test sets, while datasets associated with tasks T5 through T9 are reserved for zero-shot evaluation.
For each task, the model is trained on the combined training data from T1 to T4. For example, in T5, the model is trained using all the training sets from tasks T1 to T4 and evaluated on the T5 test set.
Notably, T8 and T9 involve entirely new diseases (asthma and pneumonia) not seen during training, allowing us to assess the model’s generalization ability to previously unseen conditions in a zero-shot setting.
Details of each dataset and task are provided in Appendix \ref{appendix: dataset-details}.

\input{tables/task_T1_T4}
\input{tables/task_T5_T9}
\subsection{Baselines and Evaluation Metrics}
\textbf{Baselines:}
We compare RespiraMFM with three state-of-the-art multimodal baselines: Qwen-2 Audio \cite{chu2024qwen2}, BTS \cite{kim2024bts} and RespLLM \cite{zhang2024respllm}. More details on baselines are included in the Appendix \ref{appendix:baselines}.

\vspace{1mm}
\noindent
\textbf{Evaluation Metrics:}
%
To ensure fair comparison, we follow prior works on respiratory disease detection to use the Area Under the Receiver Operating Characteristic Curve (AUROC) \cite{janssens2020reflection} as the evaluation metric for all the tasks.
To ensure robust evaluation, each result was obtained through three independent runs. The mean and standard deviation of the AUROC scores across these runs are reported. 

\subsection{Implementation Details}
We utilized PyTorch 2.3.0, transformers 4.47.1 \cite{wolf2020transformers}, and accelerated on four NVIDIA A100-80GB GPUs. The training process uses a batch size of 16.

%% file: tables/dataset.tex
\renewcommand{\arraystretch}{1.5}

\begin{table}[t]
    \centering
    \caption{Summary of the datasets and tasks.}
    \vspace{-2mm}
    \small
    \resizebox{\linewidth}{!}{
    \begin{tabular}{clll>{\centering\arraybackslash}p{\linewidth}}
    
        \toprule
          \textbf{Task ID}&\textbf{Dataset}& \textbf{Disease}& \textbf{\#Train/Test}\\
        \midrule
          T1&UK COVID-19 \cite{coppock2024audio}& COVID-19& 20717/11121\\
          T2&Coughvid \cite{orlandic2021coughvid}& COVID-19& 7958/2464\\
          T3&TBscreen \cite{sharma2024tbscreen}& TB&20302/8051\\
          T4&ICBHI \cite{rocha2019open}& COPD& 462/366\\
          T5&Coswara \cite{bhattacharya2023coswara}& COVID-19& -/1747\\
          T6&CodaTB \cite{huddart2024dataset}& TB& -/2053\\
          T7&KAUH \cite{fraiwan2022recognition}& COPD&-/132\\
          T8&KAUH \cite{fraiwan2022recognition}& Asthma&-/201\\
          T9&KAUH \cite{fraiwan2022recognition}& Pneumonia&-/120\\
        \bottomrule
    \end{tabular}
        }
    
    \label{tab:dataset-overview}
\end{table}






%% file: tables/task_T1_T4.tex
    
    

\renewcommand{\arraystretch}{1.5}
\begin{table*}[ht]
    \centering
    \caption{AUROC comparison for respiratory disease recognition task. Results are shown in $mean \pm std$ format of three individual runs. The \colorbox{teal!25}{teal} color indicates the best results. The values in parentheses represent the relative improvement (\%) of RespiraMFM over the strongest baseline for each task.}
    \small
    \resizebox{0.85\linewidth}{!}{
    \begin{tabular}{cl|c|c|c|c|>{\centering\arraybackslash}p{0.25\linewidth}}
    
        \toprule
        \textbf{Task ID} & \textbf{Dataset }&\textbf{Disease}&  \textbf{Qwen-2 Audio}&\textbf{BTS}& \textbf{RespLLM}& \textbf{RespiraMFM (ours)}\\
        \midrule
        T1  & UK COVID-19 &COVID-19&  0.855 $\pm$ 0.018&0.898 $\pm$ 0.010& 0.881 $\pm$ 0.005& \best 0.910 $\pm$ 0.002 (\textcolor{blue}{$\uparrow$ ~1.41 \%})\\
        T2& Coughvid &COVID-19&  0.561 $\pm$ 0.009&0.595 $\pm$ 0.014&  0.613 $\pm$ 0.011& \best 0.673 $\pm$ 0.011 (\textcolor{blue}{$\uparrow$ 9.79 \%})\\
        T3& TBscreen &TB&  0.334 $\pm$ 0.043& 0.568 $\pm$ 0.019& 0.687 $\pm$ 0.016& \best 0.709 $\pm$ 0.014 (\textcolor{blue}{$\uparrow$ 3.20 \%})\\
        T4& ICBHI &COPD&  0.614 $\pm$ 0.005&0.880 $\pm$ 0.004&  0.833 $\pm$ 0.007& \best 0.999 $\pm$ 0.000 (\textcolor{blue}{$\uparrow$ ~13.64 \%})\\
            \bottomrule
    \end{tabular}
        }
    
    \label{tab:multi-modal-train-test}
\end{table*}

%% file: tables/task_T5_T9.tex
\renewcommand{\arraystretch}{1.5}
\begin{table*}[h]
    \centering
    \caption{AUROC comparison for the respiratory disease recognition task of zero-shot prediction on new dataset. Results are shown in $mean \pm std$ format of three individual runs. The \colorbox{teal!25}{teal} color indicates the best results. The values in parentheses represent the relative improvement (\%) of RespiraMFM over the strongest baseline for each task. }
    \small
    \resizebox{0.85\linewidth}{!}{
    \begin{tabular}{cl|c|c|c|c|>{\centering\arraybackslash}p{0.25\linewidth}}
        \toprule
        \textbf{Task ID}& \textbf{Dataset }&\textbf{Task}&  \textbf{Qwen-2 Audio}&\textbf{BTS}& \textbf{RespLLM}& \textbf{RespiraMFM (ours)}\\
        \midrule
        T5  & Coswara &Covid&  0.813 $\pm$ 0.035&0.901 $\pm$ 0.008& 0.90 $\pm$ 0.006& \best 0.908 $\pm$ 0.005 (\textcolor{blue}{$\uparrow$ ~0.77 \%})\\
        T6& CodaTB &TB&   0.527 $\pm$ 0.012&0.645 $\pm$ 0.016&  0.669 $\pm$ 0.019& \best 0.689 $\pm$ 0.012 (\textcolor{blue}{$\uparrow$ ~2.99 \%})\\
        T7& KAUH &COPD&   0.581 $\pm$ 0.013&0.491 $\pm$ 0.014&  0.425 $\pm$ 0.011& \best 0.829 $\pm$ 0.005 (\textcolor{blue}{$\uparrow$ 42.74 \%})\\
        T8& KAUH &Asthma&  0.458 $\pm$ 0.010& 0.418 $\pm$ 0.016&  0.399 $\pm$ 0.010&  \best 0.552 $\pm$ 0.014 (\textcolor{blue}{$\uparrow$ ~20.55 \%})\\
        T9& KAUH &pneumonia&  0.301 $\pm$ 0.041&0.595 $\pm$ 0.020&  0.400 $\pm$ 0.021&  \best 0.709 $\pm$ 0.013 (\textcolor{blue}{$\uparrow$ 19.29 \%})\\
            \bottomrule
    \end{tabular}
        }
    \label{tab:multi-modal-zero-shot}
\end{table*}

%% file: sections/5_Results.tex
\section{Results}

\subsection{Overall Performance}
\label{RQ:overall}
First, we compare the performance of \sysname with the baselines under the supervised learning setting on the held-out test sets of the training datasets on tasks T1 through T4.
The results are summarized in Table~\ref{tab:multi-modal-train-test}. 
As shown, \texttt{RespiraMFM} consistently outperforms all other baselines across all four tasks. Overall, the average AUROC \texttt{RespiraMFM} has achieved over tasks T1 through T4 is 0.823, representing 
39.3\% improvement over Qwen2-Audio (average AUROC: 0.591), 
11.9\% improvement over BTS (average AUROC: 0.735) and 
9.15\% improvement (average AUROC: 0.754) over RespLLM. 
%
These results demonstrate the strong performance of \texttt{RespiraMFM} in identifying a wide range of respiratory diseases, advancing the state of the art.



\subsection{Zero-Shot Performance}
\label{RQ:generalize}


Next, we evaluate the zero-shot performance of \sysname under the following two scenarios.


\vspace{-2mm}
\paragraph{Unseen Datasets:} 
Regarding the unseen datasets condition, we compare \sysname with other multi-modal baselines in performing tasks T5, T6, and T7. 
In these tasks, the datasets used for evaluation are not seen during training, though the target diseases remain the same. 
Specifically, the models are trained on one or more datasets for a given disease and evaluated on a different, unseen dataset for the same condition. 
For example, in task T5, the training data includes other COVID-19 datasets such as UKCOVID-19 and CoughVid, and is evaluated on the unseen Coswara dataset.
As shown in Table~\ref{tab:multi-modal-zero-shot}, the proposed \texttt{RespiraMFM} consistently outperforms all other multi-modal baselines on these unseen datasets. 


\paragraph{Unobserved Respiratory Diseases:}
Regarding the unobserved respiratory diseases, we further compare \texttt{RespiraMFM} with \texttt{BTS} and \texttt{RespLLM} on the prediction of asthma (T8) and pneumonia (T9). In both tasks, the models are trained on datasets from T1 to T4, none of which include instances of asthma or pneumonia.
As shown in Table~\ref{tab:multi-modal-zero-shot}, despite having no disease-specific training data for these conditions, \texttt{RespiraMFM} consistently outperforms all the baselines. 
Specifically, \texttt{RespiraMFM} achieves a 20.55\% and 19.29\% relative improvement in asthma and pneumonia prediction over the other baselines, respectively. 
Our average AUROC over these tasks (T5-T9) is 0.738, outperforming 
Qwen2-Audio by 37.69\% (average 0.54), 
\texttt{BTS} by 20.98\% (average 0.61) and 
\texttt{RespLLM} by 32.02\% (average 0.56) on average AUROC.
Overall, these results suggest that \texttt{RespiraMFM} generalizes effectively across datasets and to previously unseen respiratory diseases.


\subsection{Effects of Data Scaling}

\label{RQ:data-model-scaling}
\begin{figure}[]
    \centering
    \begin{subfigure}[b]{\columnwidth} 
        \centering
        \includegraphics[width=\textwidth]{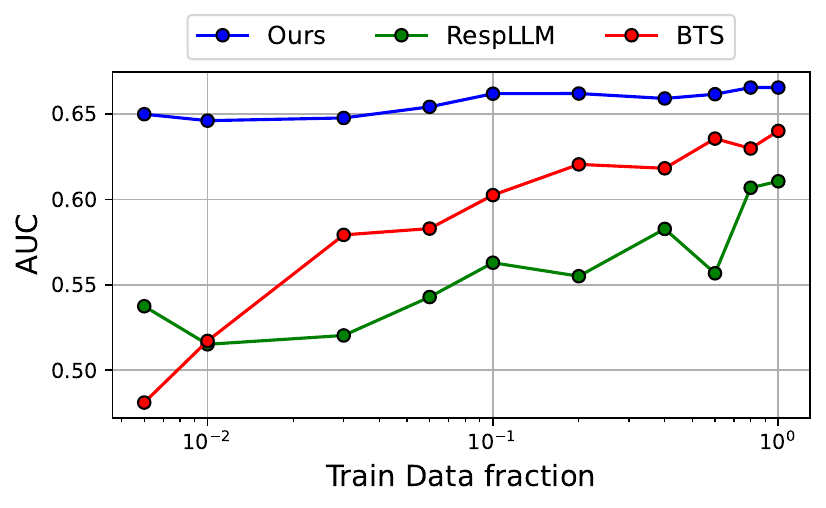} 
        \caption{Single-Modal}
           \vspace{7mm}
        \label{fig:data-scaling (a)}
    \end{subfigure}
 
    \hfill
    \begin{subfigure}[b]{0.95\columnwidth}
        \centering
        \includegraphics[width=\textwidth]{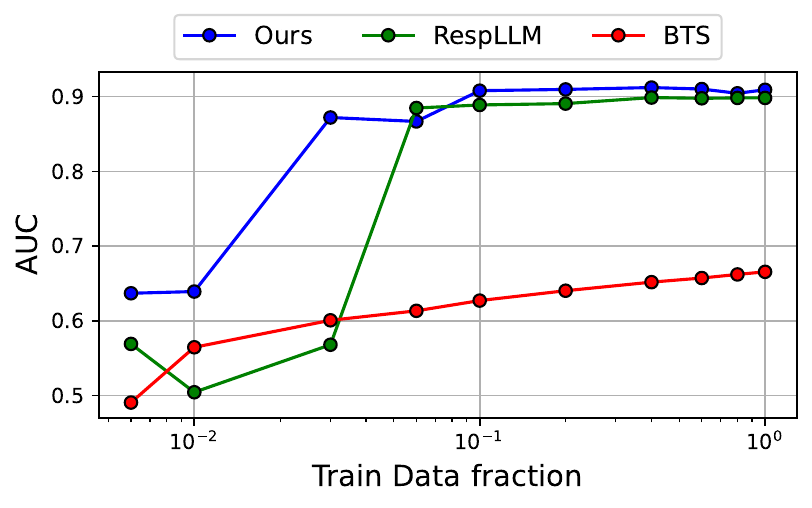} 
        \caption{Multimodal}
        \label{fig:data-scaling (b)}
    \end{subfigure}
    \caption{Effect of dataset scaling. }
    \label{fig:data-scaling}
\end{figure}

\noindent
To assess how the training dataset size impacts the model performance, we conducted experiments on Task T1 by systematically varying the number of training examples. 
In this experiment, the model was trained on the UKCOVID-19 dataset for 1 epoch, and evaluated on the held-out test of the UKCOVID-19 dataset. Starting with a full training set, we randomly sampled subsets at varying fractions and compared our model with the baselines on the same test set. 
We explored two configurations for this experiment: a single-modal setup using only audio features as input, and a multi-modal setup that integrates both audio and textual features as input.
The results are shown in Figure \ref{fig:data-scaling}. Figure \ref{fig:data-scaling (a)}, which corresponds to the single-modal setting using only audio input, shows a clear trend of improved performance with increasing training samples, indicating that larger datasets lead to better performance. Our model consistently outperforms both BTS and RespLLM across all data fractions, with notably strong performance even at low data availability. While all models benefit from more data, ours maintains a consistent lead.
In contrast, Figure \ref{fig:data-scaling (b)} illustrates the multi-modal configuration, where both audio and text features are used as input. Here, our model rapidly approaches peak performance with minimal training data and significantly outperforms the baselines across nearly all data scales. These results highlight the strength of multi-modal integration, especially in clinical contexts where labeled data is often limited. The findings suggest that multi-modal models are particularly well-suited for deployment in resource-constrained healthcare settings, offering high diagnostic performance even with sparse training data.
%
Even in the audio-only setting shown in Figure \ref{fig:data-scaling (a)}, the contrastive alignment still plays a role, just not during inference. During training, the contrastive projection head is trained on the available audio–text pairs using the same fraction of training data. This pretraining step aligns the audio representations with the corresponding symptom text, improving the structure of the learned audio embedding space and therefore providing a more realistic initialization for fine-tuning the downstream tasks, even in the unimodal experiment. During evaluation, however, we use audio-only input from the test set, without providing any text.
This means that although the model receives only audio at inference time, it benefits from having learned a better-aligned embedding space during contrastive pretraining. Thus, the model maintains stronger audio-only performance, and the gains persist even when symptoms are missing.

\subsection{Ablation Study}
\label{RQ:multi-modal}
\vspace{1mm}

\vspace{1mm}
\noindent
\input{tables/multimodal}
\noindent
\textbf{Uni-Modality vs. Multi-Modality}:
To assess the effectiveness of multimodal integration compared to unimodal inputs, we conducted experiments on Task T5, aiming to understand whether combining audio and textual information offers complementary benefits that improve diagnostic performance beyond what a single modality can achieve alone.
In this experiment, the model is trained on the combined data from all available training datasets and evaluated in a zero-shot setting on the Coswara dataset. We select the Coswara dataset for this experiment because it provides both disease labels and additional metadata describing patient health status, including severity levels such as asymptomatic (no symptoms), mild, moderate, and healthy.
We group these into three broad categories—mild or no symptoms, moderate symptoms, and healthy—and evaluate models in three configurations: audio-only input (uni-modal), text-only input (uni-modal), and multimodal input combining both audio and text. Accuracy is used as the evaluation metric for all configurations.
As shown in Table \ref{tab:multi-modal}, for cases with mild or no symptoms, the audio-only model outperforms the text-only model based on the symptom information. Conversely, the text-only model performs better compared to the audio-only model for symptomatic and healthy individuals.
On the other hand, the multimodal model, which integrates both audio and text information, consistently outperforms both unimodal models across all severity levels and on the overall dataset. In summary, these results demonstrate the clear advantage of combining multiple modalities for improved disease prediction.



\begin{figure}[t]
    \centering
    \includegraphics[width=\linewidth]{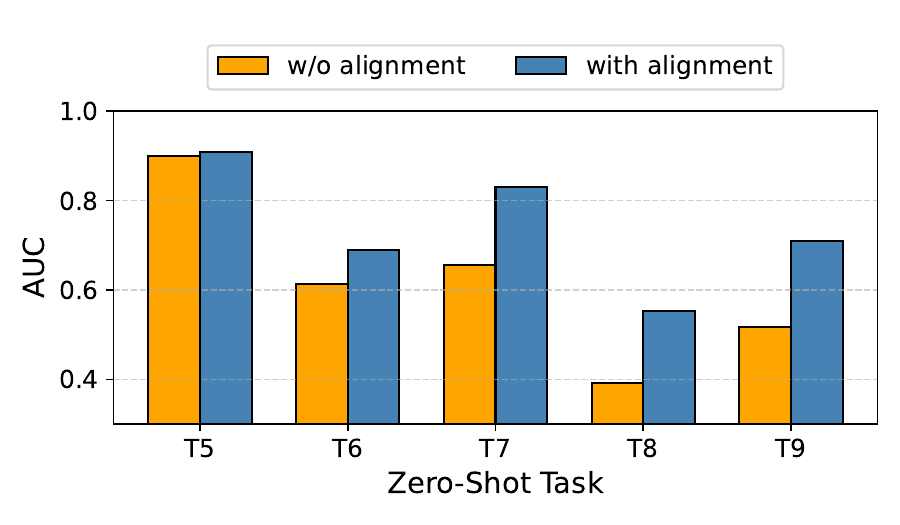}
    \caption{AUC of zero-shot disease detection task with and without alignment module}
    \label{fig:radar_aligner_projector}
\end{figure}

\begin{figure}[t]
    \centering
    \begin{subfigure}[b]{0.49\columnwidth}
        \centering
        \includegraphics[width=\textwidth]{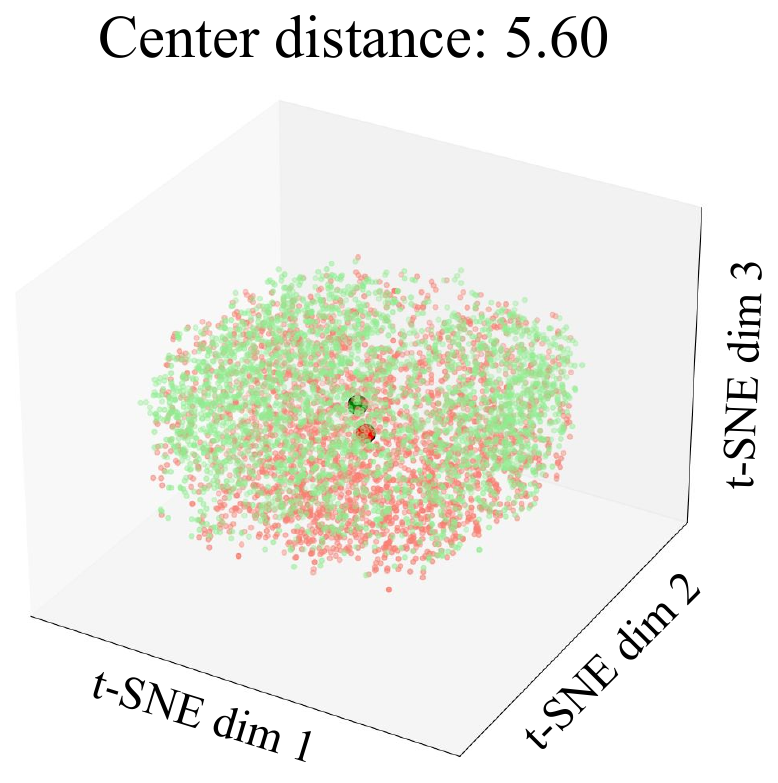} 
        \caption{without alignment}
        \label{fig:tsne-alignment (a)}
    \end{subfigure}
    \hfill
    \begin{subfigure}[b]{0.49\columnwidth}
        \centering
        \includegraphics[width=\textwidth]{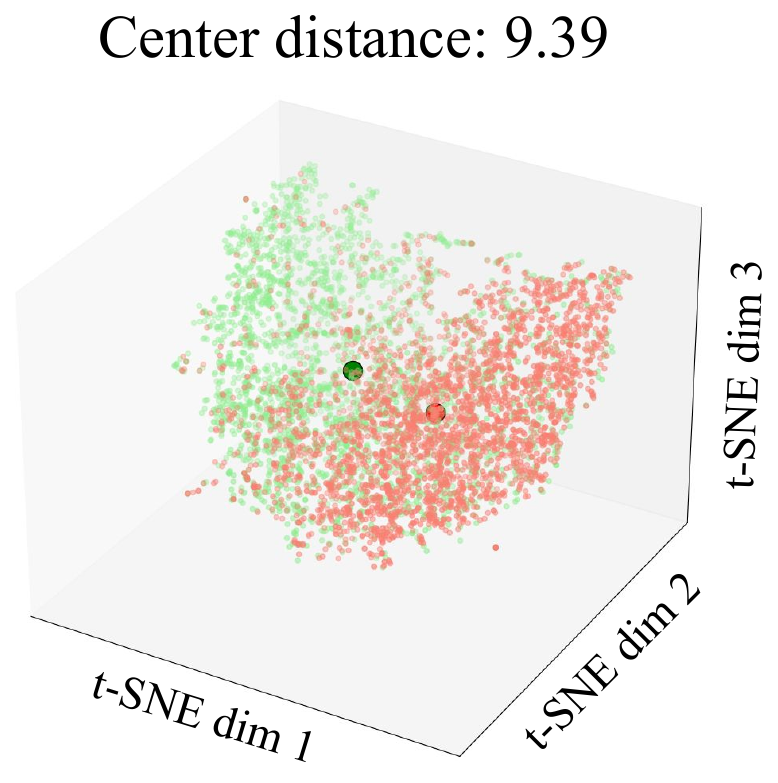} 
        \caption{with alignment}
        \label{fig:tsne-alignment(b)}
    \end{subfigure}
    \caption{t-SNE visualization of audio embeddings for ukcovid-19 dataset. The left panel shows audio embeddings obtained without alignment, while the right panel shows those with alignment. The \textcolor{green}{green} points represent sample from \textcolor{green}{healthy} patients and \textcolor{red}{red} points represent sample from \textcolor{red}{covid} patients.}
    \label{fig:tsne-alignment}
    \vspace{3mm}
\end{figure}

\vspace{1mm}
\noindent
\textbf{Analysis on alignment module}:
\label{ablation:alignment-process}
\noindent
To evaluate the effectiveness of the contrastive alignment module introduced in \S\ref{section:training-projector}, we conducted experiments across all the zero-shot tasks (T5-T9) by training the model both with and without this component.
For the baseline setting (without alignment), we used a standard linear projector and trained it jointly with the LLM in a single-stage fine-tuning process. In contrast, for the alignment setting, we first trained the projector using a contrastive loss and then kept it frozen during LLM fine-tuning.
The results are summarized in Figure~\ref{fig:radar_aligner_projector}. As shown, the alignment module consistently improves AUC scores across all tasks, suggesting that it provides more informative feature representations for instruction tuning.
To further understand why the improvement on zero-shot tasks occurs, we analyzed the learned audio embeddings from the projectors and visualized them using 3D t-SNE (Figure \ref{fig:tsne-alignment}). The left panel (\ref{fig:tsne-alignment (a)}) shows embeddings obtained without alignment, while the right panel (\ref{fig:tsne-alignment(b)}) shows those with alignment. Samples from COVID patients are colored in red, and healthy samples are colored in green. The embeddings with alignment form clearer clusters and are more separable between the two classes compared to the baseline. 
This enhanced separability allows the model to learn more discriminative and semantically aligned audio features, thereby improving generalization and leading to better zero-shot performance.

    
    

\vspace{1mm}
\noindent
\textbf{Interpretability analysis}:
\label{ablation:alignment-process}
\begin{figure}[t]
    \centering
    \begin{subfigure}[b]{0.5\columnwidth} 
        \centering
        \includegraphics[width=\textwidth]{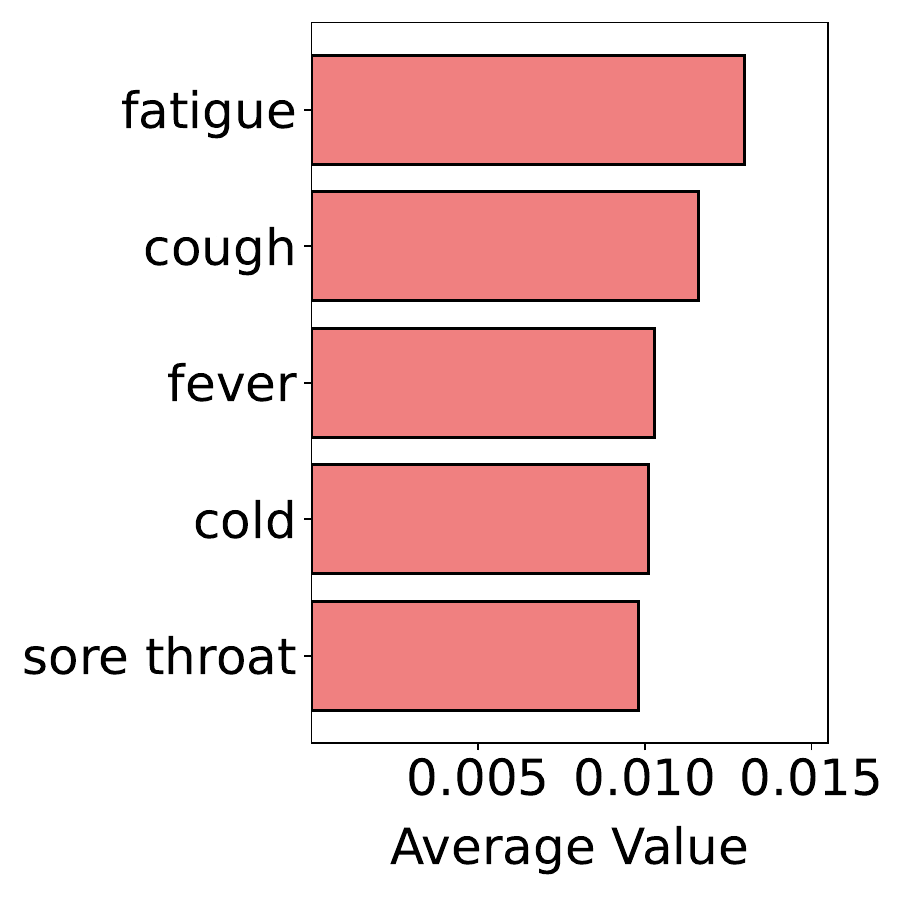} 
        \caption{Covid Patients}
        \label{fig:top5-attn-map(a)}
    \end{subfigure}
    \hfill
    \begin{subfigure}[b]{0.43\columnwidth}
        \centering
        \includegraphics[width=\textwidth]{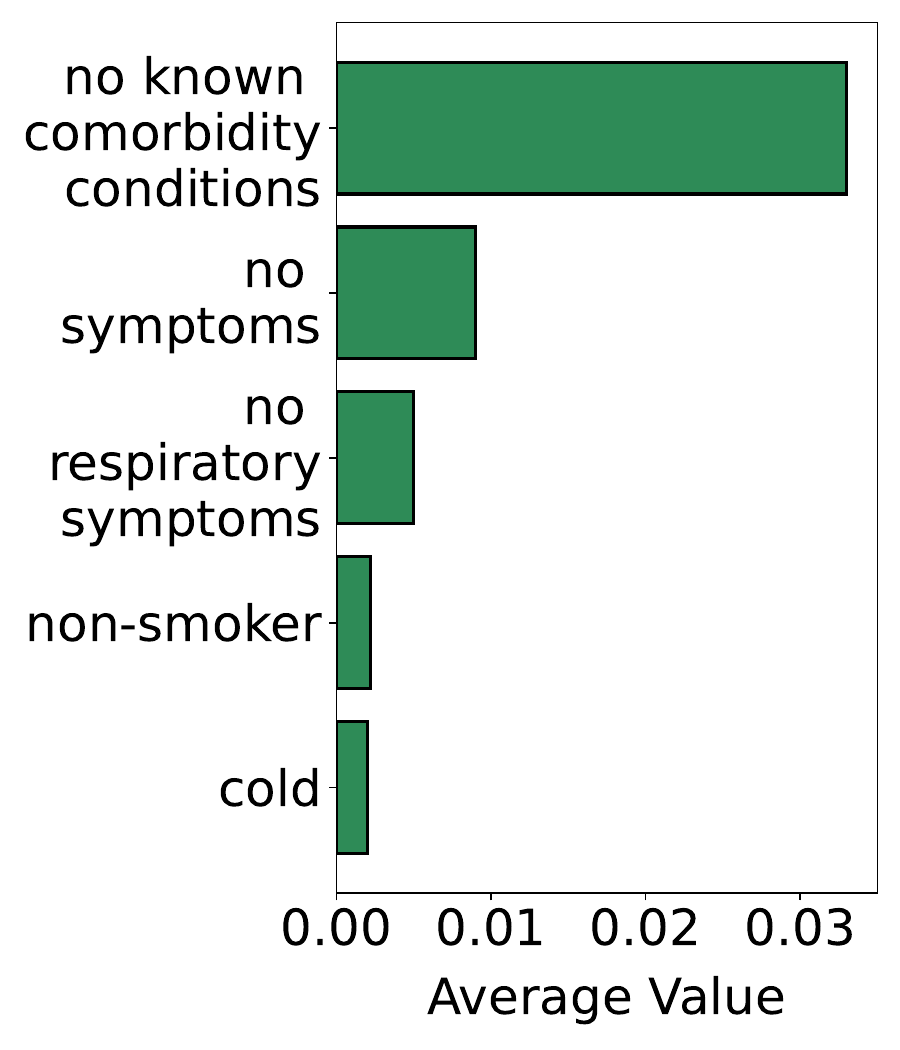} 
        \caption{Healthy Patients}
        \label{fig:top5-attn-map(b)}
    \end{subfigure}
    \caption{Average attention weight of top-5 tokens for both healthy and COVID-19 patients on the Coswara dataset.}
    \vspace{3mm}
    \label{fig:top5-attn-map}
\end{figure}
\noindent
To evaluate the interpretability of our models, we extracted the \texttt{[CLS]} token attention and measured the average attention weight assigned to each context token. As shown in Figure \ref{fig:top5-attn-map}, we conducted experiments separately for healthy and COVID-positive patients and report the average attention across their respective tokens or groups of tokens. From the plot, we can see that for COVID-positive patients, most of the attention is concentrated on tokens representing fever, cough, fatigue, and other symptomatic indicators, which is clinically expected. On the other hand, for healthy patients, most of the attention is placed on token groups that explicitly represent the absence of symptoms (e.g., no symptoms, no respiratory symptoms, non-smoker). This indicates the model not only attends to disease-related symptoms when present but also leverages the absence of symptoms for decision-making. 
These observations highlight that the model’s focus aligns with medically relevant patterns, indicating a degree of interpretability in an otherwise black-box architecture.


%% file: tables/multimodal.tex
\begin{table}
    \centering
    \resizebox{\columnwidth}{!}{
    \begin{tabular}{ccccc}
    \toprule
         &  \textbf{\makecell{Mild or No\\symptoms}}&  \textbf{\makecell{Moderate\\symptoms}}&  \textbf{Healthy}&  \textbf{Total}\\
         \midrule
         Audio&  \underline{0.3576}&  0.3571&  0.7266&  0.6102\\
         Text&  0.3294&  \underline{0.619} &  \underline{0.9766}&  \underline{0.7934}\\
         Audio+Text&  \textbf{0.4047}&  \textbf{0.6587}&  \textbf{0.9849}&  \textbf{0.8203}\\
         \bottomrule
    \end{tabular}
    }
    \caption{Performance comparison of audio-only, text-only, and multimodal (audio+text) models across different patient groups in the Coswara dataset. \textbf{Bold} indicates the best performance and \underline{underlined} indicates the second-best.}
    \label{tab:multi-modal}
\end{table}

%% file: sections/6_Conclusion.tex
\section{Conclusion}
In this paper, we introduced RespiraMFM, a multimodal foundation model designed to detect respiratory diseases by integrating respiratory sound recordings with patient-reported symptoms and medical history. 
We proposed an effective method for multimodal alignment of text and audio input, demonstrating strong performance across nine tasks involving five major respiratory diseases using diverse real-world datasets. 
We also showed that the model can maintain high diagnostic accuracy even with limited training data, making it suitable for deployment in data-scarce healthcare environments. 
Overall, RespiraMFM offers a scalable, non-invasive, and clinically relevant solution for early and accurate respiratory disease detection, with the potential to support medical professionals and improve decision-making across a variety of healthcare settings.

%% file: sections/7_Limitations.tex
\section{Limitation}
While our proposed multimodal foundation model shows strong performance across various respiratory disease detection tasks, it has some limitations. The model's effectiveness depends on the quality and consistency of symptom metadata, which can differ significantly between datasets and clinical environments. 
Furthermore, the availability of evaluation data is uneven across diseases. Datasets for COPD, asthma, and pneumonia (Tasks T7, T8, T9) contain substantially fewer samples than those for COVID-19 and tuberculosis, resulting in small test set sizes that may limit the statistical reliability of performance estimates for these conditions. Expanding evaluation to larger and more balanced datasets for these underrepresented diseases would strengthen the generalizability claims of the framework.
Additionally, although the model integrates audio and symptom data, incorporating additional modalities such as medical imaging or wearable sensor data could further improve its diagnostic accuracy and robustness.

%% file: sections/8_Ethical.tex
\section{Ethics Statement}
We foresee no ethical concerns with our work. All the datasets used in this study were anonymized and excluded any participant identity information.

\section{Acknowledgment}
We sincerely thank the reviewers and ACs for their constructive comments. 
Shakhrul Iman Siam and Mi Zhang were supported in part by NSF Award NeTS-2312675. Tiantian Feng and Shrikanth Narayanan were supported by funds from NSF and NIH.

%% file: sections/Appendix.tex
\appendix
\twocolumn

\section{Additional Details on Datasets}
\label{appendix: dataset-details}
In this study, we used the following datasets:

\vspace{1mm}
\noindent
\textbf{UK COVID-19}:
The UK COVID-19 Vocal Audio Dataset \cite{coppock2024audio} represents the largest collection of SARS-CoV-2 PCR-referenced audio recordings to date, compiled in the United Kingdom. 
The dataset features PCR test results linked to 70,794 out of 72,999 participants, with 24,155 of the 25,776 confirmed positive cases accurately documented. Notably, respiratory symptoms were reported by 45.62\% of the participants, providing valuable symptomatic metadata for analysis.
All the audio recordings were captured in the .wav format.  In our study, we adopt the official train-test split released with the dataset.

\input{plots/dataset_stats}

\begin{figure}[h!]
    \centering
    \begin{subfigure}[b]{\columnwidth} 
        \centering
        \includegraphics[width=\columnwidth]{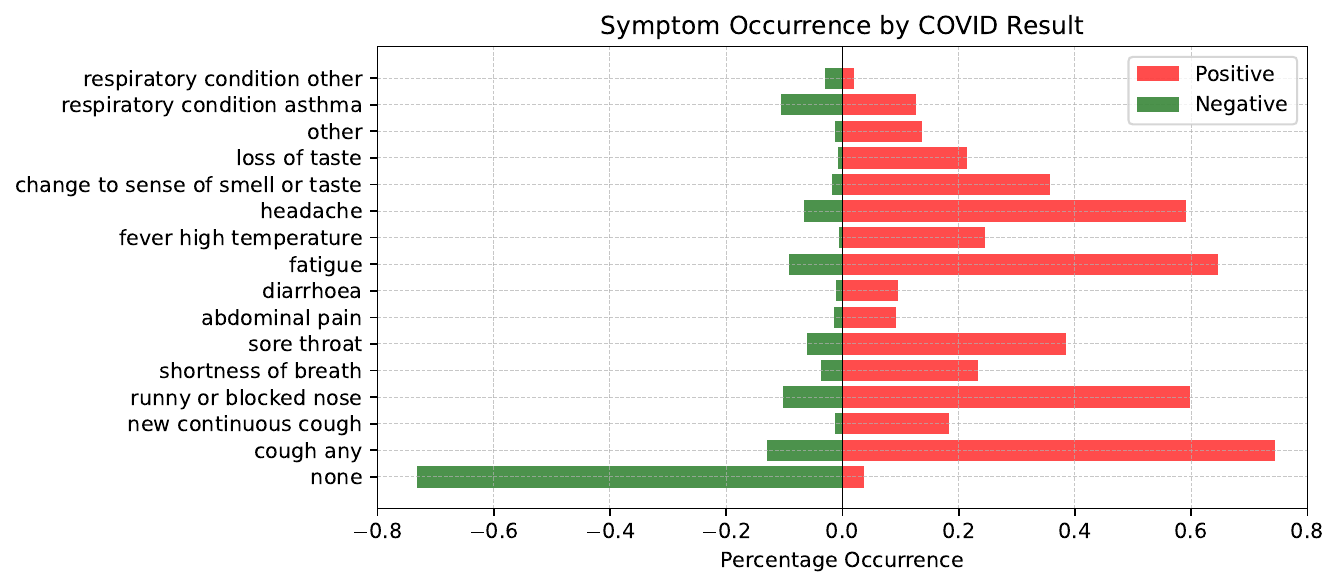} 
        \caption{UK-covid19 dataset}
        \label{fig:learning-rate}
    \end{subfigure}
    \hfill
    \begin{subfigure}[b]{\columnwidth}
        \centering
        \includegraphics[width=\columnwidth]{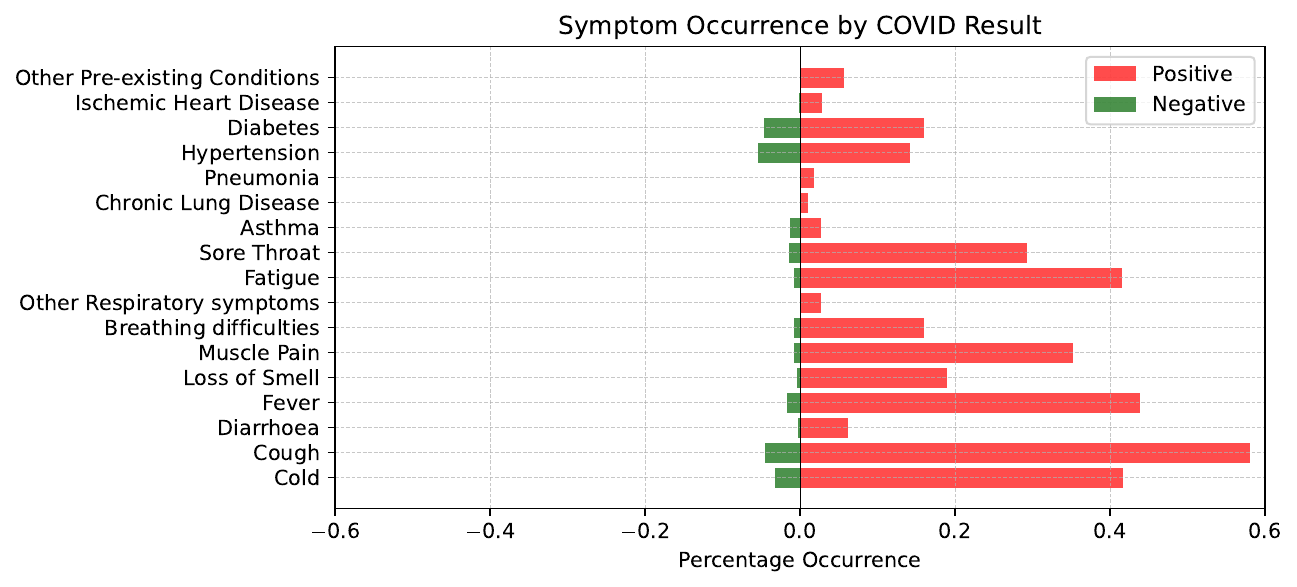} 
        \caption{Coswara dataset}
        \label{fig:training-loss-LR}
    \end{subfigure}
    \caption{Symptom Occurrence Distribution by COVID-19 Test Result in UK COVID-19 and Coswara Datasets.}
    \label{fig:1x2images}
\end{figure}

\vspace{1mm}
\noindent
\textbf{Coswara}:
The Coswara dataset \cite{bhattacharya2023coswara} is a diverse collection of respiratory sounds and detailed metadata, recorded between April 2020 and February 2022 from 2,635 individuals, including 1,819 SARS-CoV-2 negative, 674 positive, and 142 recovered cases. It features nine categories of respiratory sounds, covering variations of breathing, coughing, and speech, providing a rich dataset for analyzing respiratory health. In addition to audio recordings, the dataset includes comprehensive metadata, capturing demographic details such as age, gender, and geographic location, along with health-related information like symptoms, pre-existing respiratory conditions, comorbidities, and COVID-19 test status. We follow the official data split, which contains 70\% samples for training, 15\% for validation, and 15\% for testing.

\vspace{1mm}
\noindent
\textbf{COUGHVID}:
The COUGHVID dataset \cite{orlandic2021coughvid} is a large-scale, publicly available collection of over 25,000 crowdsourced cough recordings, covering a diverse range of ages, genders, geographic locations, and COVID-19 statuses. The database contains approximately 35 hours of audio recordings, comprising around 37,000 segmented cough samples. An automatic cough classifier was used to filter recordings, retaining only those with a minimum probability of 0.8 of containing cough sounds. The final distribution of labeled recordings was as follows: 25\% COVID-positive cases, 35\% symptomatic cases, 25\% healthy individuals, and 15\% with no reported health status.

\vspace{1mm}
\noindent
\textbf{TBscreen}:
The TBscreen dataset \cite{sharma2024tbscreen} was collected in Nairobi and comprises cough recordings from 149 subjects diagnosed with pulmonary tuberculosis (TB) and 46 control subjects with other respiratory illnesses. The dataset includes a total of 33,000 passive coughs and 1,600 forced coughs, all recorded in a controlled setting to ensure consistency across subjects with similar demographics. To standardize the data for applications, each cough recording was processed to have a fixed duration of one second. Longer recordings were segmented into multiple one-second audio files, while shorter recordings were centered and padded with zeros to maintain uniformity.

\vspace{1mm}
\noindent
\textbf{CodaTB}:
The CodaTB dataset \cite{huddart2024dataset} is a large, multi-country collection of cough sounds from individuals undergoing evaluation for tuberculosis (TB). It comprises over 700,000 cough recordings from 2,143 participants, along with detailed demographic, clinical, and microbiological diagnostic information. The dataset was collected as part of broader TB research studies, where participants underwent a baseline questionnaire, clinical examination, and sputum collection for TB testing at the time of enrollment. Comprehensive metadata accompanies the cough recordings, including age, gender, height, weight, smoking status, and duration of cough. Additionally, HIV status was determined either through self-reported diagnosis or confirmed positive test results. The dataset was split into training (n = 1,105) and validation (n = 1,038) subsets.

\vspace{1mm}
\noindent
\textbf{ICBHI}:
The ICBHI Respiratory Sound Database \cite{rocha2019open} was originally compiled to support the International Conference on Biomedical Health Informatics (ICBHI) 2017 scientific challenge and is now publicly available for research. It consists of a combination of public and private datasets collected independently by two research teams across two different countries over several years. The dataset contains 5.5 hours of respiratory sound recordings, comprising 6,898 respiratory cycles from 126 subjects. The 920 audio samples in the dataset have been manually annotated by respiratory experts, classifying them based on the presence of crackles, wheezes, both, or no adventitious respiratory sounds. Additionally, the dataset provides diagnostic labels for chronic obstructive pulmonary disease (COPD), pneumonia, and asthma, enabling the development of machine-learning models for disease classification.

\vspace{1mm}
\noindent
\textbf{KAUH}:
The KAUH (King Abdulaziz University Hospital) dataset \cite{fraiwan2022recognition} is a collection of respiratory sound recordings from 112 subjects, including 35 healthy individuals and 77 patients with pulmonary conditions. 
Lung sounds were recorded using an electronic stethoscope, which was placed at multiple points on the chest wall to capture respiratory sounds while avoiding heart sounds. The recordings were processed and extracted using Heart and Lung Sound Visualization software, which allows exporting data with three different filter settings (Bell, Diaphragm, and Extended) to emphasize different frequency ranges relevant to lung sounds.

\input{tables/symptom_table}

\section{More Details of Audio Encoder}
We utilized the Opera-CT encoder \cite{zhang2024opera}, to extract audio features from raw audio signals. Opera-CT is a contrastive learning-based hierarchical token-semantic audio transformer \cite{chen2022htsat}. It operates by dividing the mel-spectrogram into patches, which are embedded as input tokens for the transformer. The model leverages a hierarchical architecture with window attention, optimizing both computational efficiency and memory usage by restricting attention to localized windows. The transformer has 31 million parameters and produces output features of size $D_a = 768$.




\section{Baselines}
\label{appendix:baselines}
We compared \sysname with the following state-of-the-art multimodal baselines:

\vspace{1mm}
\noindent
\textbf{Qwen2-Audio}:
Qwen2-Audio \cite{chu2024qwen2} is a large-scale multimodal language model capable of processing both speech and text inputs. It leverages the Qwen2 backbone with an integrated audio encoder that converts raw waveforms into latent representations aligned with textual embeddings through a shared multimodal transformer. For our experiments, we utilized the publicly released Qwen2-Audio model \footnote{https://github.com/QwenLM/Qwen2-Audio}.
We implemented an additional MLP classification head and fine-tuned the model on our training set in order to obtain continuous prediction scores and compute AUROC fairly across all models.

\vspace{1mm}
\noindent
\textbf{BTS}:
BTS \cite{kim2024bts} proposes a module called Bridging Text and Sound (BTS), which aligns respiratory audio and text metadata by utilizing CLAP \cite{elizalde2023clap} as a dual-purpose encoder for both modalities. In this approach, CLAP independently processes text and audio data through separate encoders. The resulting embeddings are then concatenated and passed through a linear classifier to perform the disease prediction.

\vspace{1mm}
\noindent
\textbf{RespLLM}:
RespLLM \cite{zhang2024respllm} introduces a multimodal approach using a pre-trained audio encoder and a Large Language Model for diagnosing respiratory diseases using audio recordings and patient metadata. RespLLM employs a trainable linear projector to align audio embeddings with the language model's input space. In contrast, our method adopts a contrastively trained projection head, which enables more effective alignment between audio and text modalities.

\section{Additional Details on Contrastive Aligner}
\label{appendix: contrastive-aligner}
\subsection{Model Architecture}
The contrastive alignment module is implemented as a multi-layer perceptron (MLP) with normalization and regularization components. Specifically, the projection head maps an input embedding of dimension 768 into a higher-dimensional contrastive space of $D_{LLM}$ through an intermediate hidden layer of size 1024. The architecture consists of a linear transformation followed by Layer Normalization, ReLU activation, and dropout (rate = 0.1). A final linear layer produces the output embeddings used for contrastive supervision.

\subsection{Training}
We trained the alignment module using the same dataset employed during instruction-tuning. The model was optimized for 500 epochs with a learning rate of 0.001.
We used a standard contrastive loss temperature of 0.07.



\section{Additional Details on Instruction Tuning}
Table \ref{tab:hyperparams} presents the hyperparameter settings used in this work.

\begin{table}[h]
\centering
\begin{tabular}{lc}
    \midrule[1.2pt]
    \textbf{Hyperparameters} & \textbf{Value} \\
    \midrule
    Instruction tuning epochs & 20 \\
    LoRA alpha & 32 \\
    LoRA rank & 16 \\
    LoRA dropout & 0.1 \\
    Total batch size & 16 \\
    Maximum sequence length & 256 \\
    Learning rate & 1e-5 \\
    Learning rate optimizer & AdamW \\
    Schedule & linear \\
    Weight decay & 0.1 \\
    \midrule[1.2pt]
\end{tabular}
\caption{Training hyperparameters}
\label{tab:hyperparams}
\end{table}


\section{Additional Experiments}

\input{tables/model_architecture}
\noindent
\textbf{Ablation on Model Architecture}:
Table \ref{tab:ablation-model-architecture} presents the results across all tasks using various LLMs as a backbone. Among them, Phi-2 consistently outperforms other models, achieving the highest average score across all tasks.
This outcome aligns with prior findings from the “Textbooks Are All You Need” \cite{gunasekar2023textbooks}, and “The surprising power of small language models” \cite{javaheripi2023phi}, which demonstrate that the 2.7B parameter Phi-2 model can outperform significantly larger models, including 7B and 13B architectures across multiple reasoning benchmarks. This counterintuitive result is largely attributable to Phi-2’s high-quality, curriculum-style training data, which includes synthetic datasets explicitly designed to teach common-sense reasoning, scientific knowledge, and biomedical facts. Moreover, Phi-2 benefits from a knowledge-scaling pipeline, where a carefully trained 1.3B model (Phi-1.5) is distilled and expanded into the 2.7B Phi-2, enabling efficient knowledge transfer and strong downstream performance.
Our observations in Table \ref{tab:ablation-model-architecture} are consistent with these findings: Phi-2, despite its smaller size, achieves better results than the more general-purpose LLaMA-3 8B in our setting. 

\begin{figure}[t]
    \centering
    \includegraphics[width=\linewidth]{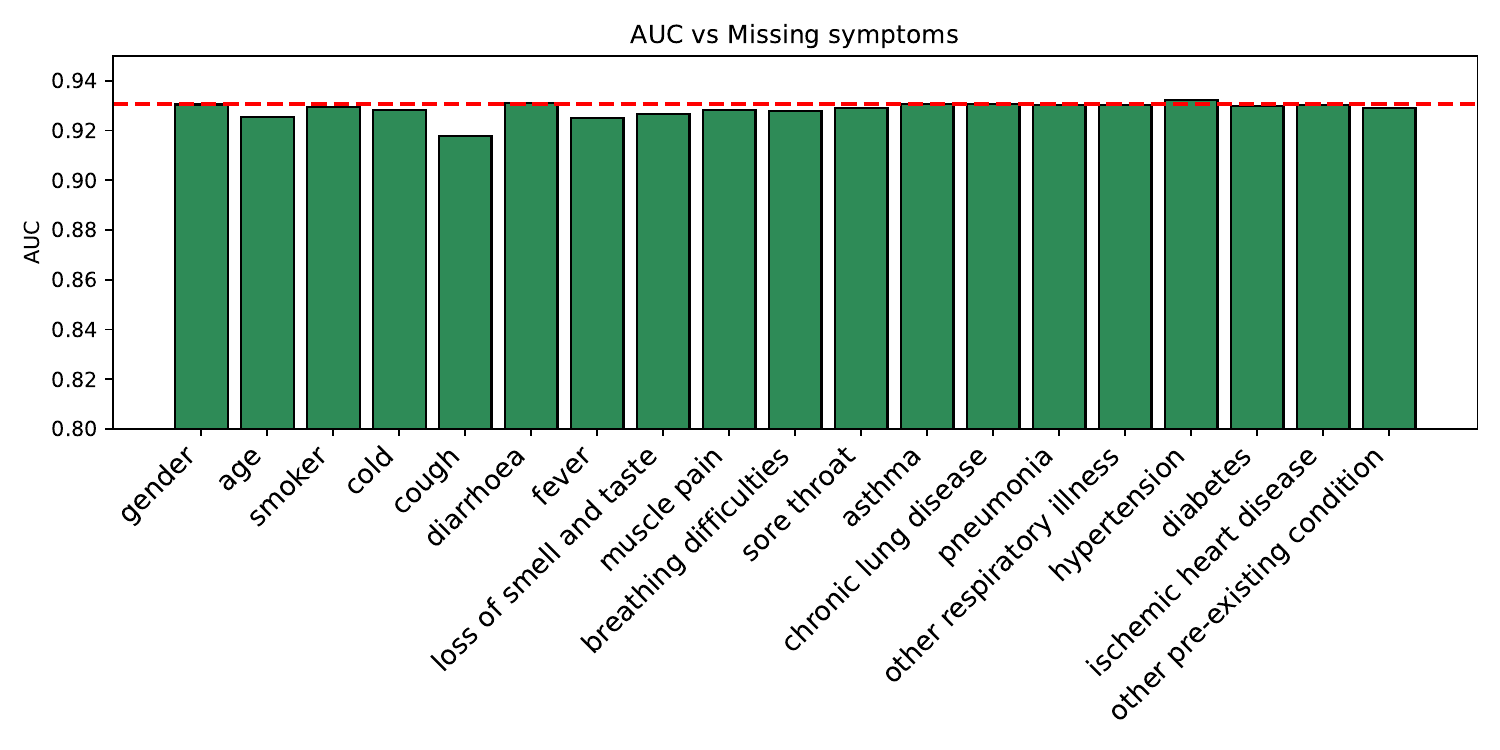}
        \caption{AUC on Task T5 under various missing input scenerios.The red dotted line represents the AUC when all symptoms are present.}
    \label{fig:missing-symptoms}
\end{figure}

\vspace{1mm}
\noindent
\textbf{Missing Input}:
\noindent
To simulate a real-world scenario where some data may be missing, we evaluate \sysname under various missing input conditions, as shown in Figure \ref{fig:missing-symptoms}. This experiment is performed on task T5. Specifically, we remove one symptom from the patient metadata, restructure the prompt, and feed it to the model to calculate the AUC. The red dotted line represents the AUC when all symptoms are present, serving as a baseline. As observed in the plot, removing information such as gender does not significantly affect performance, with AUC either remaining the same or showing only a slight decrease. However, the removal of symptoms like having cough results in a more noticeable drop in AUC compared to other symptoms, suggesting that these symptoms play a more critical role, which aligns with expectations in COVID detection. Although the AUC decreases when certain information is removed, the drop is not substantial, indicating that the system is sufficiently robust to handle missing inputs in real-world scenarios.

\begin{figure}[t]
    \centering
    \includegraphics[width=\linewidth]{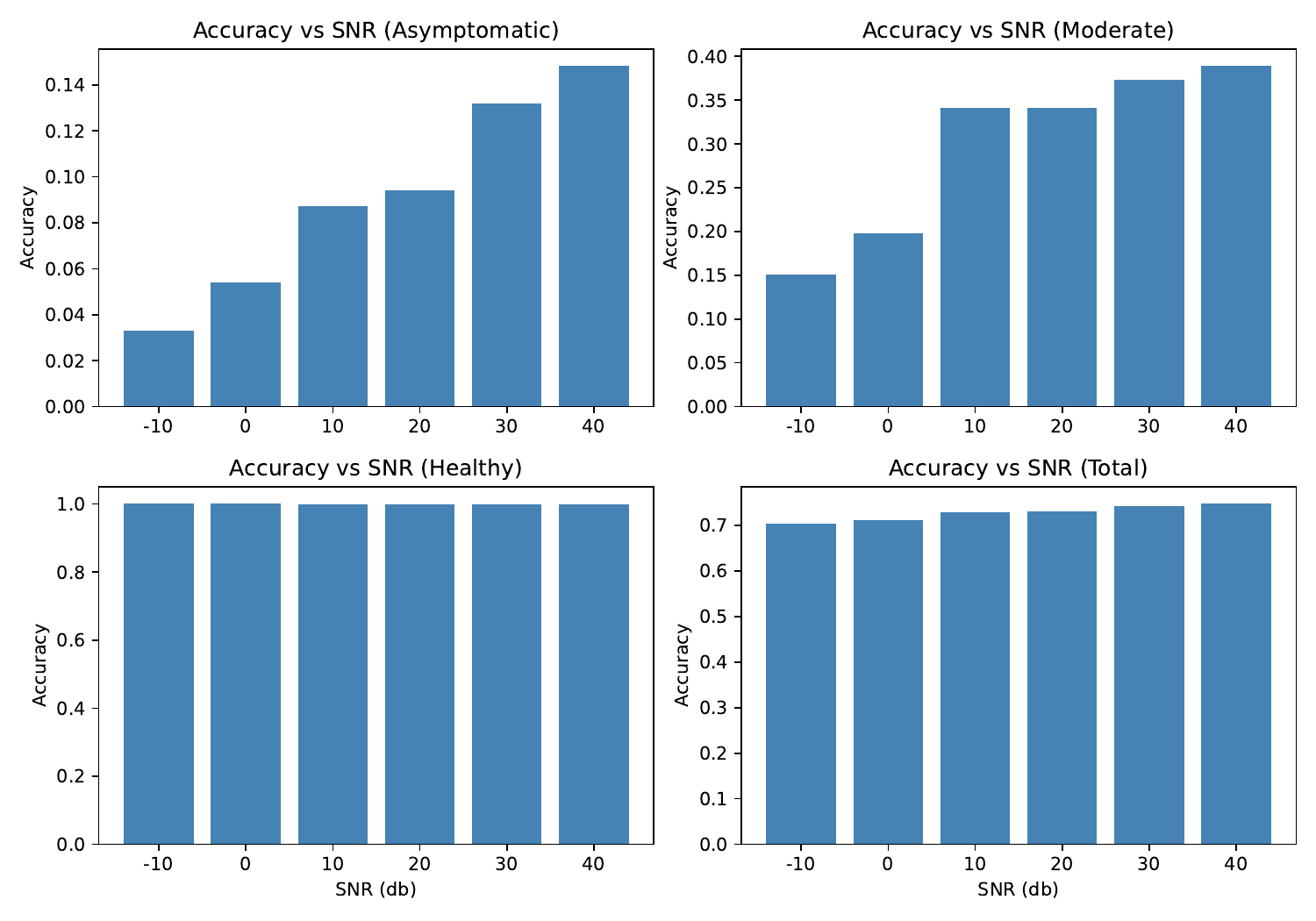}
        \caption{AUC on Task T5 under noisy condition.}
    \label{fig:noisy-input}
\end{figure}

\vspace{1mm}
\noindent
\textbf{Experiment on Noisy Input}: To evaluate the robustness of \sysname under noisy conditions, we conducted experiments by adding Gaussian noise to the audio inputs. We selected Task 5, corresponding to zero-shot evaluation on the Coswara dataset, as it provides both disease labels and detailed patient metadata, including health status and severity levels such as asymptomatic, moderate, and healthy. The results, shown in Figure \ref{fig:noisy-input}, illustrate the model’s performance across these conditions. We observe that under the asymptomatic setting where subtle acoustic cues are most critical, adding noise noticeably degrades performance. In contrast, for the healthy group, the addition of noise has minimal impact, indicating that the model remains relatively robust in scenarios with less disease-related acoustic variability.

\begin{figure}[t]
    \centering
    \includegraphics[width=0.7\linewidth]{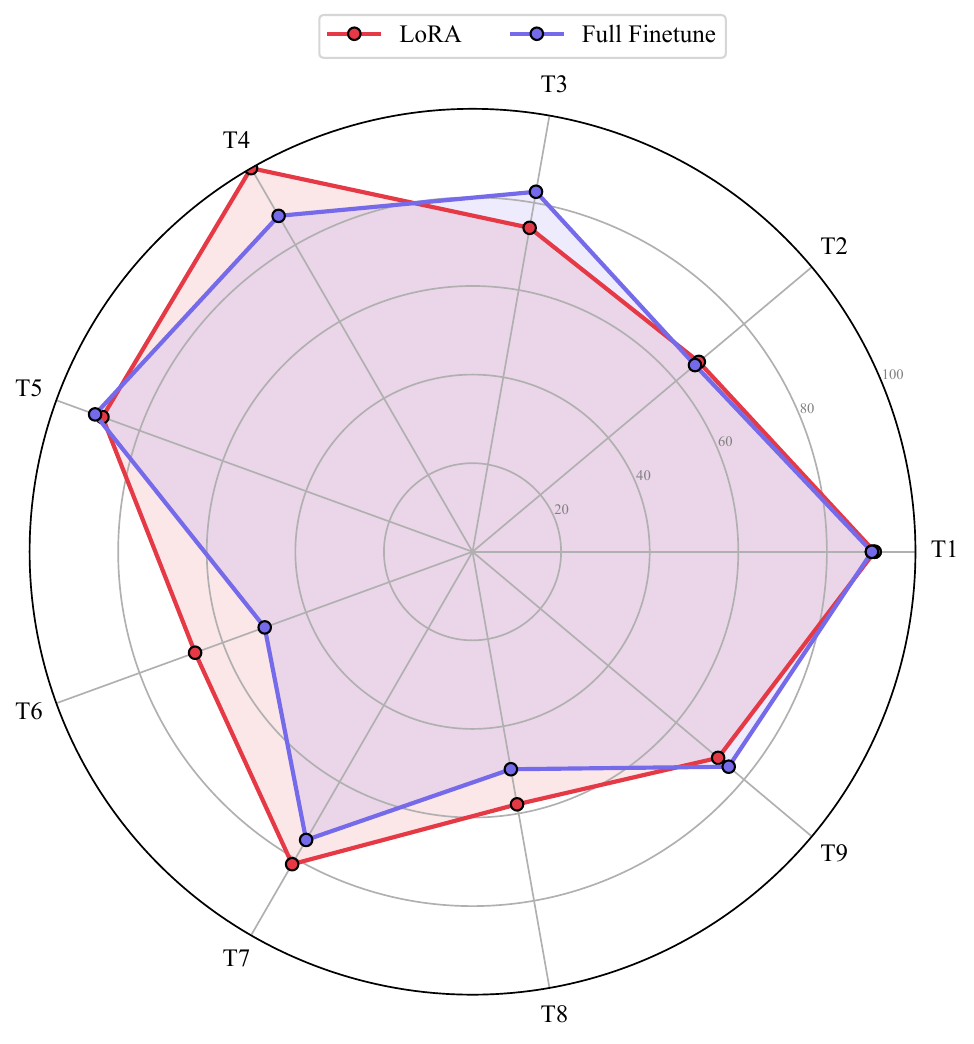}
        \caption{LoRA vs Full Fine-tuning AUC comparison.}
    \label{fig:lora-vs-full-radar}
\end{figure}

\noindent
\textbf{LoRA vs.\ Full Fine-Tuning of the LLM Backbone:}
\label{app:lora-vs-full}
To justify our choice of low-rank adaptation (LoRA) for the Phi-2 backbone, we compare it against full fine-tuning under an identical optimization budget. Both variants use the same OPERA-CT audio encoder and contrastive projector, and share all training hyperparameters.
The only difference is whether the backbone weights are updated in full or through LoRA adapters. 
Figure~\ref{fig:lora-vs-full-radar} visualizes per-task AUC. LoRA improves on six of nine tasks, with the largest margins on T6 ($+0.167$), T4 ($+0.124$), T8 ($+0.080$), and T7 ($+0.063$); full fine-tuning is competitive only on T3, T5, and T9.
As these results come from a single run per setting without variance estimates, we treat the comparison as indicative rather than conclusive, but it consistently favors the parameter-efficient configuration used in the main paper. 

\begin{figure}[t]
    \centering
    \includegraphics[width=0.7\linewidth]{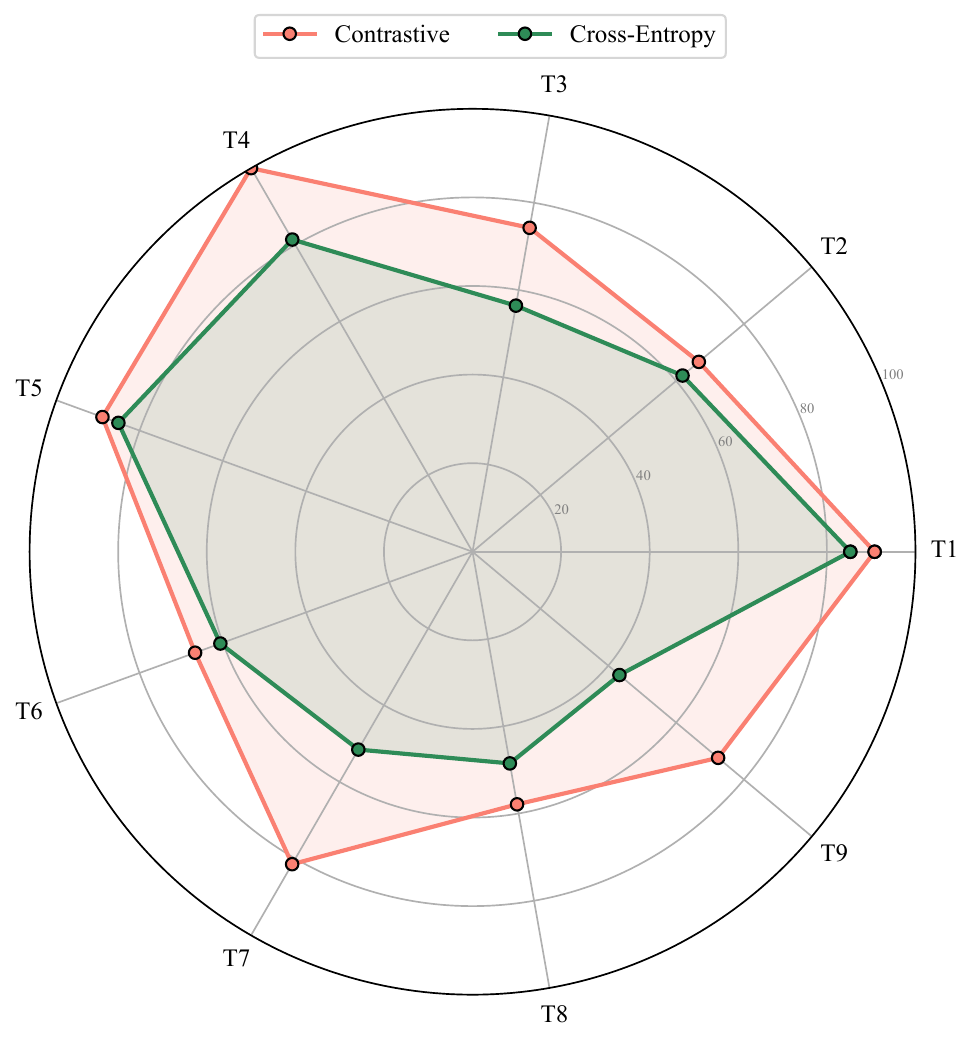}
    \caption{Per-task comparison of stage-1 projector objectives.}
    \label{fig:proj-objective-radar}
\end{figure}

\noindent
\textbf{Projector Training Objective: Contrastive vs.\ Cross-Entropy}
We ablate the objective used to train the projector in stage~1, which maps OPERA-CT audio embeddings into the language model's input space. We compare two variants. \emph{Contrastive} aligns each audio embedding with its paired text description through an audio-text contrastive loss, as used in the main paper. \emph{Cross-Entropy (CE)} instead trains the projector directly with a classification loss on the ground-truth task labels, without any text supervision. All other components, data, and downstream stage-2 settings are held fixed; only the stage-1 objective differs.
Figure~\ref{fig:proj-objective-radar} reports per-task performance across the nine evaluation tasks (T1--T9). The contrastive objective matches or exceeds CE on every task, with the largest margins on T3, T4, T7, and T8, while the two are comparable only on T1 and T5. The advantage is consistent rather than driven by any single task. These results support our use of audio-text contrastive pre-training as the
default stage-1 objective.

%% file: plots/dataset_stats.tex
\begin{figure}[htp]
    \begin{subfigure}[b]{\columnwidth}
        \centering
        \includegraphics[width=\columnwidth]{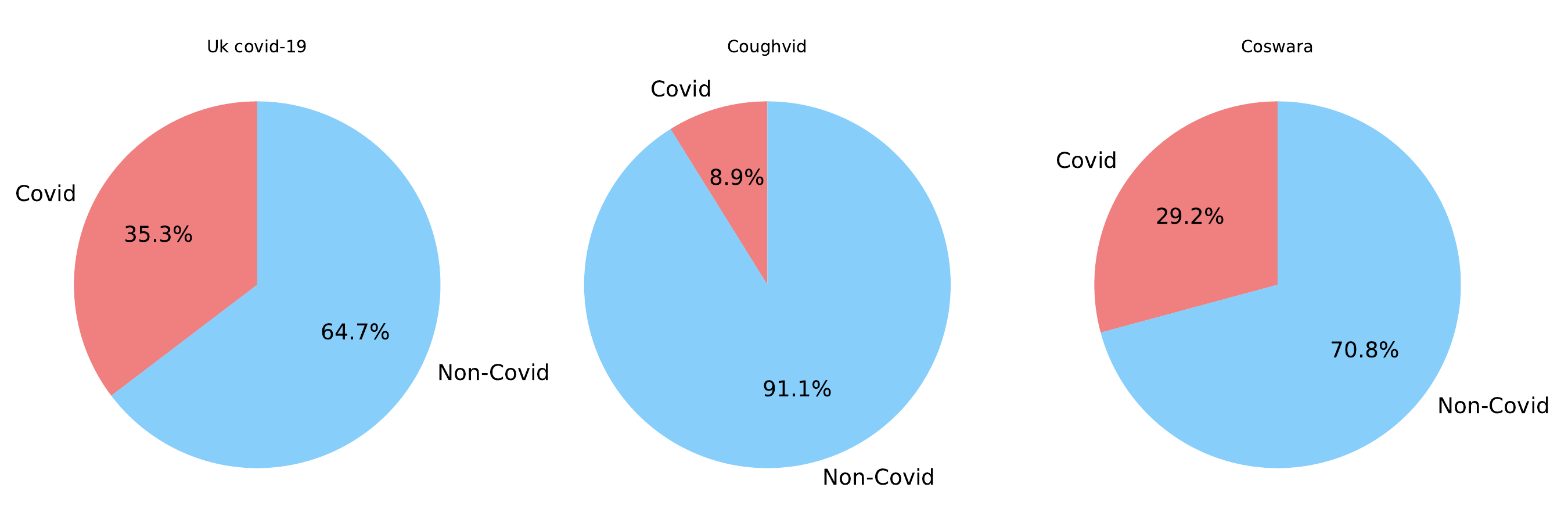} 
        \caption{Class Distribution in covid datasets (UK covid-19, coughvid and coswara)}
        \label{fig:dataset1}
    \end{subfigure}
    
    \vspace{10pt} 

    \begin{subfigure}[b]{\columnwidth}
        \centering
        \includegraphics[width=\columnwidth]{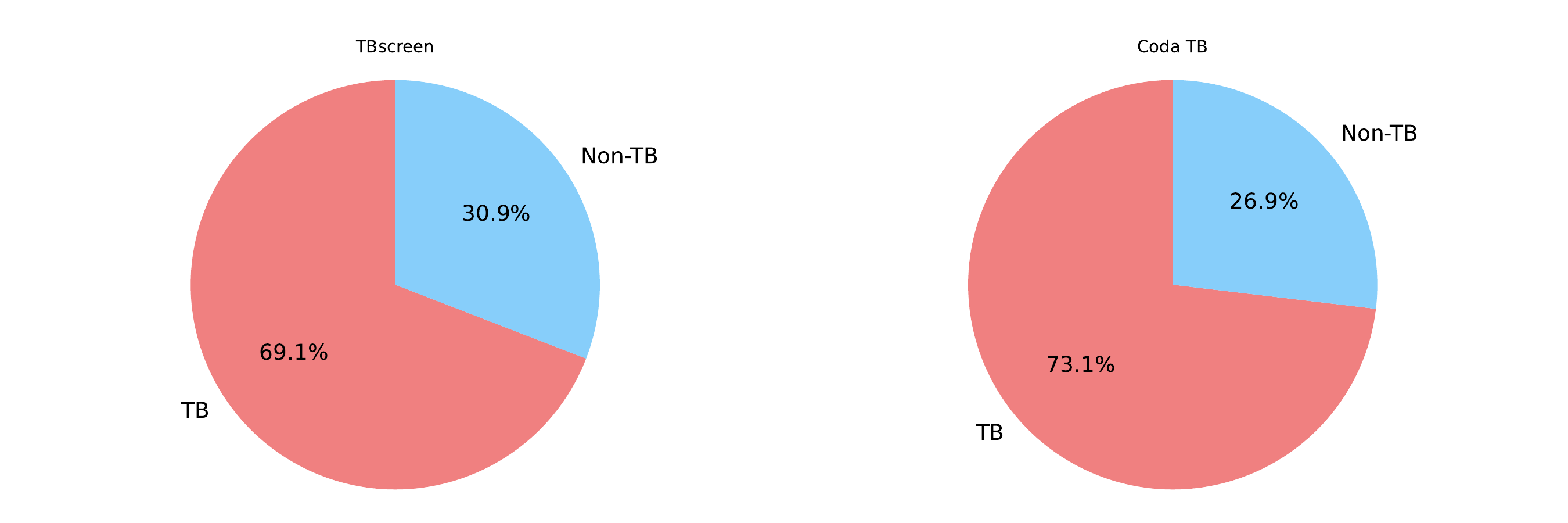} 
        \caption{Class Distribution in TB datasets (TBscreen and Coda TB)}
        \label{fig:dataset2}
    \end{subfigure}
    
    \vspace{10pt} 

    \begin{subfigure}[b]{\columnwidth}
        \centering
        \includegraphics[width=\columnwidth]{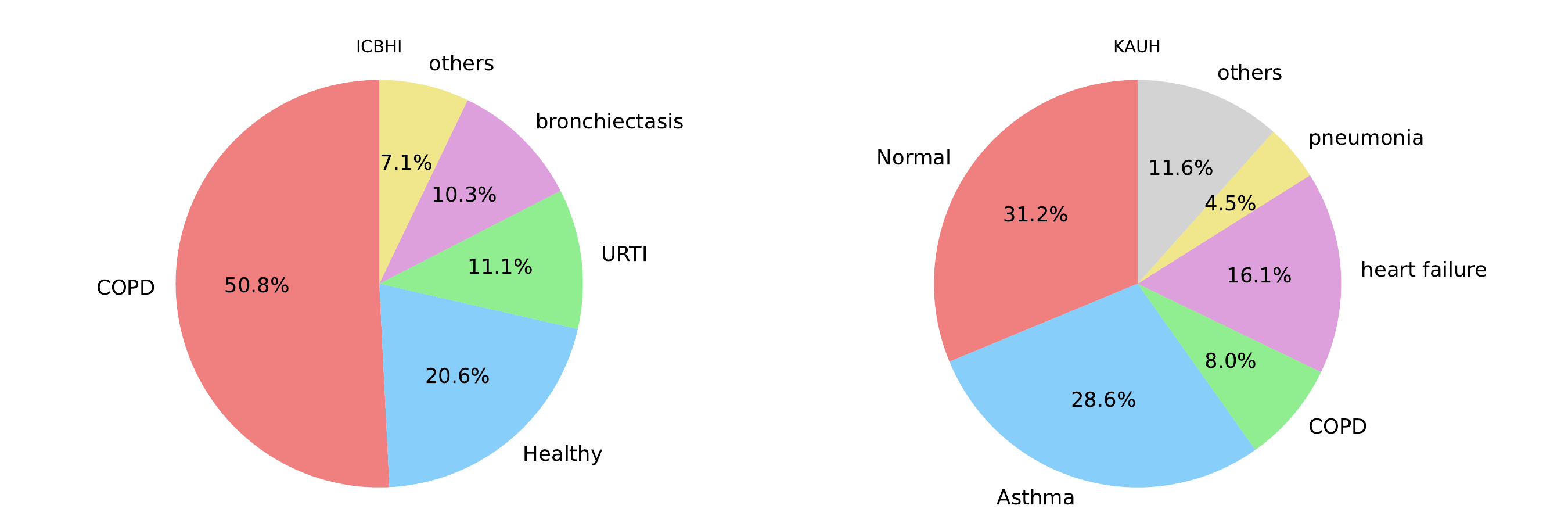} 
        \caption{Class Distribution in ICBHI and KAUH datasets}
        \label{fig:dataset3}
    \end{subfigure}
    
    \caption{Class Distribution Across Datasets}
    \label{fig:classification_distribution}
\end{figure}

%% file: tables/symptom_table.tex
\renewcommand{\arraystretch}{1.3}

\begin{table*}[h]
    \centering
    \caption{Dataset-wise patient symptoms and medical history selection}
    \resizebox{\textwidth}{!}{
    \begin{tabular}{>{\centering\arraybackslash}p{0.25\linewidth}|>{\raggedright\arraybackslash}p{0.86\linewidth}}
    
        \toprule
         \textbf{Dataset}& \textbf{Patient Information}\\
        \midrule
         UK COVID-19&  Age, sex, cough, new continuous cough, runny or blocked nose, shortness of breath, sore throat, abdominal pain, diarrhea, fatigue, fever, headache, changes to sense of smell or taste, loss of taste, asthma, other symptoms  \\ 
         COUGHVID & Age, sex, fever and muscle pain, other respiratory symptoms \\
         TBscreen& Age, sex, fever, cough, night sweats, cough with blood, smoking status, previous TB history, HIV status, cough duration\\
         ICBHI & Age, sex, BMI, child weight, child height, recording device placement\\
         Coswara& Age, sex, cold, cough, diarrhea, fever, loss of smell and taste, muscle pain, breathing difficulties, fatigue, sore throat\\
         CodaTB& Age, sex, fever, weight loss, night sweats, cough with blood, previous TB history, HIV status, cough duration\\
         KAUH& Age, sex, recording device placement, sound type\\
            \bottomrule
    \end{tabular}
        }
    
    \label{tab:selected-symptoms}
\end{table*}

%% file: tables/model_architecture.tex
\begin{table*}[h]
    \centering
    \caption{AUROC comparison for all respiratory disease recognition tasks using different LLMs as backbone. Here `M’ denotes the million level, and `B’ denotes the billion level. The
    \colorbox{teal!40}{heavy teal} color indicates the highest results.}
    \scalebox{0.80}{
    \begin{tabular}{c|cc|ccccccccc|c}
    \toprule[1.2pt]
          \textsc{Models}&\textsc{Size} &\textsc{Dim}&  \textbf{T1}&  \textbf{T2}&  \textbf{T3}&  \textbf{T4}&  \textbf{T5}&  \textbf{T6}&  \textbf{T7}&  \textbf{T8}& \textbf{T9} &\textbf{Average}\\
    \midrule[1.2pt]
         GPT2-Medium &345M    &1024&  \best 0.911 & \best 0.688 & 0.773 & 0.826 & \best 0.918 & 0.508 & 0.713 & \best 0.605& 0.677 & 0.735\\
         LLaMA-3 &1B          &2048&  0.906 & 0.645 & 0.767 & 0.996 & 0.913 & 0.563 & 0.686 & 0.547 & 0.619 & 0.738\\
         Phi-2 &2.7B          &2560&  0.910& 0.673&  0.709& \best 0.999& 0.907& 0.689& \best 0.829&  0.552& \best 0.709& \best 0.776\\
         LLaMA-3&8B &4096&  0.905 & 0.676 & \best 0.796& 0.851 & 0.853 & \best 0.711 & 0.642 & 0.594 & 0.617 & 0.738\\
    \bottomrule
    \end{tabular}
    }
    \label{tab:ablation-model-architecture}
\end{table*}